\documentclass[fleqn,usenatbib]{mnras}

\usepackage{newtxtext,newtxmath}

\usepackage[T1]{fontenc}

\DeclareRobustCommand{\VAN}[3]{#2}
\let\VANthebibliography\thebibliography
\def\thebibliography{\DeclareRobustCommand{\VAN}[3]{##3}\VANthebibliography}

\usepackage{graphicx}	
\usepackage{amsmath}	
\usepackage{physics}

\newcommand{\HI}{{\sc H\,i}\ }

\title[Chemistry during cloud formation]{Time-dependent chemical evolution during cloud formation: H$_2$-regulated chemistry in diffuse molecular cloud}
\author[Y. Komichi et al.]{
Yuto Komichi,$^{1}$\thanks{E-mail: komichi-yuto832@g.ecc.u-tokyo.ac.jp}
Yuri Aikawa,$^{1}$
Kazunari Iwasaki$^{2}$
and Kenji Furuya$^{3}$
\\
$^{1}$Department of Astronomy, Graduate School of Science, The University of Tokyo, 7-3-1, Hongo, Bunkyo-ku, Tokyo 113-0033, Japan\\
$^{2}$Center for Computational Astrophysics, National Astronomical Observatory of Japan, 2-21-1, Osawa, Mitaka, Tokyo 181-8588, Japan\\
$^{2}$RIKEN Pioneering Research Institute, 2-1 Hirosawa, Wako-shi, Saitama 351-0198, Japan
}

\date{Accepted XXX. Received YYY; in original form ZZZ}

\pubyear{\the\year{}}

\begin{document}
\label{firstpage}
\pagerange{\pageref{firstpage}--\pageref{lastpage}}
\maketitle

\begin{abstract}

We investigate the chemical evolution of a forming molecular cloud behind an interstellar shock wave. We conduct three-dimensional magnetohydrodynamics simulations of the converging flow of atomic gas, including a simple chemical network and tracer particles that move along the local velocity field. Then we perform detailed chemical network calculations along the trajectory of each tracer particle. The diffuse part of forming molecular clouds is CO-poor; i.e., H$_2$ and CO abundances do not correlate. In diffuse regions of $n_\mathrm{H}\lesssim 10^{3}\,\mathrm{cm^{-3}}$, we find that the abundances of hydrocarbons and oxygen-bearing molecules are determined by steady-state chemistry reflecting the local H$_2$ abundance, which is determined by the gas density along the trajectory. In denser regions, the abundances are affected by water ice formation, which changes the elemental abundance of carbon and oxygen (i.e., C/O ratio) in the gas phase. Assuming quasi-steady-state chemistry given the abundances of major molecules (e.g., H$_2$) from the simple network, we derive analytic solutions for molecular abundances, which reproduce the calculation results. We also calculate the molecular column densities based on the spatial distribution of tracer particles and their molecular abundances, and compare them with observations of diffuse molecular clouds. We find that the column densities of CH, CCH, and OH are linearly correlated with those of H$_2$, which supports the empirical relation used in the observations. On the other hand, the column density of HCO$^+$ shows non-linear dependence on the H$_2$ column density, reflecting the difference in HCO$^+$ formation paths in CO-poor and CO-rich regions. 


\end{abstract}




\section{Introduction}\label{sec:introduction}

The formation of molecular clouds is a crucial process of astrophysics that gives initial conditions for star formation. Molecular cloud formation is thought to be triggered by interstellar shock waves driven by astronomical energetic events such as supernovae and massive star formations \citep{Inutsuka_2015, Pineda_2023}; compression and accumulation of atomic gas by shock waves at multiple times result in dense gas. Recent theoretical works show that this framework may explain various characteristics of molecular clouds, such as the origin of filamentary structures \citep{Inoue_2013, Inoue_2018, Abe_2021, Gomez_2014, Gomez_2018, Zamora-Aviles_2018} and the mass functions of giant molecular clouds \citep{Kobayashi_2017, Kobayashi_2018}. Recent observations of bubble structures in nearby galaxies by James Webb Space Telescope (JWST) also suggest the importance of stellar feedback effects on molecular clouds and star formation \citep{Watkins_2023}. 

The formation phase of molecular clouds is also important for astrochemistry. As the cloud formation proceeds, atomic H is converted to H$_2$, which is the most abundant molecule in the interstellar medium (ISM). Because of the low-density and low-pressure environment, H$_2$ is mainly formed through grain surface reactions. The timescale of H$_2$ formation is given by that of collisions between H atoms and dust grains: $\sim 10\,\mathrm{Myr}\,(n/100\,\mathrm{cm^{-3}})^{-1}$, where $n$ is gas number density. This is comparable to the timescale of gas accumulation by interstellar shock waves: $\sim 6\,\mathrm{Myr}\,(A_V/1\,\mathrm{mag})(v_\mathrm{flow}/10\,\mathrm{km\,s^{-1}})^{-1}(n_\mathrm{flow}/10\,\mathrm{cm^{-3}})^{-1}$ \citep{Hartmann_2001,Pringle_2001,Inoue_2009}, where $A_V$ is the visual extinction of accumulated gas, $v_\mathrm{flow}$ is the inflowing gas velocity, and $n_\mathrm{flow}$ is the inflowing gas number density. It indicates that H$_2$ formation is non-equilibrium, i.e., H$_2$ formation is affected by the physical evolution of the ISM. 

The abundances of molecules other than H$_2$ have been mainly studied by detailed chemical network calculations. Most of them assume static physical conditions; e.g., pseudo-time-dependent model \citep[e.g.][]{Prasad_1980b,Harada_2019} and the photodissociation region model \citep[e.g.][]{Tielens_1985}. The effect of turbulent diffusion on molecular abundances in static interstellar clouds is also investigated \citep[e.g.][]{Xie_1995}. However, the physical evolution of the ISM can also affect these molecules, since their formation rates depend on the H$_2$ abundance. In this work, we investigate the formation processes of interstellar molecules especially the effect of non-equilibrium H$_2$ chemistry during molecular cloud formation by combining chemical network calculations and magnetohydrodynamics simulations.

The cloud formation processes are studied with simulations of shock compressions or colliding flows \citep{Koyama_2002, Heitsch_2006, Heitsch_2009, Hennebelle_2007, Inoue_2008, Inoue_2009, Inoue_2012, Clark_2012, Clark_2019, 1977ApJ...218..148M, Iwasaki_2019, Iwasaki_2022, Komichi_2024}. Some of them solved a simple chemical network and hydrodynamics simultaneously, not only to treat detailed gas cooling processes but also to study the chemical properties during the transition from \HI gas to molecular clouds. \citet{Valdivia_2016} focused on H$_2$ formation during shock compression of atomic gas. They found that H$_2$ is rapidly formed within dense clumps in a compression layer and is transferred to lower-density regions via turbulent mixing. It results in abundant H$_2$ in the diffuse medium, which cannot be achieved solely by the H$_2$ formation in the diffuse gas because of its long timescale. \citet{Clark_2012, Clark_2019} included CO formation and found that CO formation is slow compared with H$_2$ formation. They conducted radiation transfer calculations and confirmed that CO traces only dense parts of molecular clouds. Similar findings are suggested by larger-scale calculations focusing on the local galactic plane \citep{Seifried_2020, Seifried_2022, Bellomi_2020} and the simulations of interstellar medium in turbulence \citep[e.g.][]{Glover_2010}.

Although the previous studies revealed the basic pictures of major molecules such as H$_2$ and CO during the cloud formation, the effects of gas dynamics on lower-abundance molecules are still not understood. In the early stage of molecular cloud formation phase, the typical number density and visual extinction are  $n\sim 10-10^3 \,\mathrm{cm^{-3}}$ and $A_V \sim 0.1 - 3\,\mathrm{mag}$, respectively. These quantities are consistent with diffuse and translucent molecular clouds \citep{Snow_2006}. Thus, they are suitable targets to compare with our results. One of the remaining problems of diffuse and translucent molecular clouds is the spatial distribution of H$_2$ and other molecules. Since the energy difference between the ground and first excited state is about $170\,\mathrm{K}$, the rotation of H$_2$ is not excited in low temperature conditions ($T< 100\,\mathrm{K}$). Although H$_2$ in molecular clouds can be observed with absorption lines, such lines are detected only along limited lines of sight because of their short wavelength. Another way to investigate H$_2$ distribution is to observe CO, but the spatial distributions of H$_2$ and CO are not necessarily the same due to their different formation conditions. This means there should be so-called "CO-dark" molecular clouds, which cannot be probed by CO emission or absorption \citep[e.g.][]{vanDishoeck_1992}. 

Therefore, diffuse and translucent molecular clouds in our Galaxy are often observed with absorption lines of various molecules toward light sources such as stars, extragalactic sources, and so on. Previous studies derived column densities of molecules such as hydrocarbons and cyanides and found correlations between them, although there is a scatter of orders of magnitude among the clouds \citep[e.g.][]{Liszt_1996,Lucas_1996,Lucas_2000,Godard_2010,Gerin_2011,Neufeld_2015,Jacob_2022,Kim_2023}. Some molecules, such as CH, are used as H$_2$ tracers \citep{Liszt_2002,Sheffer_2008}. We aim to provide theoretical explanations of such empirical abundance correlations.

Because of limited computational resources, including detailed chemical networks in the multi-dimensional hydrodynamics simulations is not feasible. 
To reduce the computational cost, previous work adopted reduced chemical networks for specific molecules \citep[e.g.][]{Xu_2019, Grassi_2022}, steady-state chemistry \citep[e.g.][]{Bisbas_2012, Valdivia_2017}, or autoencoders to predict molecular abundances \citep[e.g.][]{Holdship_2021}. 
However, \citet{Komichi_2024} pointed out that ice formation on grain surfaces affects the molecular abundance in the gas phase even in the molecular cloud formation phase. Since the grain surface chemistry is basically not in a steady-state and includes hundreds of important reactions, there is a risk in adopting the methodologies above. Alternatively, we can use tracer particles for the detailed chemical network calculations \citep[e.g.][]{vanWeeren_2009, Furuya_2012, Jensen_2021, Panessa_2023, Priestley_2023}. We can trace the trajectories of fluid elements and record the time evolution of physical quantities. Conducting chemical network calculations along the trajectories of tracer particles allows us to look into the effect of gas dynamics and its history on chemical evolution. A similar approach is taken with smoothed particle hydrodynamics simulations \citep[e.g.][]{Wakelam_2019, Clement_2023}. Since we aim to analyze the relationship between gas dynamics and detailed molecular evolution during molecular cloud formation, we adopted this approach in magnetohydrodynamics simulations of shock compressions of atomic gas in this work.

The rest of this paper is organized as follows. The methods of MHD simulations and chemical network calculations with tracer particles are summarized in Section \ref{sec:method}. The results of the MHD simulation and chemical network calculations are shown in Section \ref{sec:Results}. We analyze and compare our results with observations of diffuse and translucent clouds in Section \ref{sec:discussion}. Finally, we summarize the present work in Section \ref{sec:conclusion}.

\section{Methods}\label{sec:method}

\subsection{MHD simulation}\label{sec:MHD}

\subsubsection{Basic physics}\label{sec:MHDequ}

The basic equations are given as follows;

\begin{equation}
    \pdv{\rho}{t}+\nabla\cdot (\rho \vb*{v})=0,
\end{equation}
\begin{equation}
        \pdv{t}(\rho\vb*{v})+\boldsymbol\nabla\cdot\biggl[\rho\vb*{v}\boldsymbol\otimes\vb*{v}+\left(p+\frac{\vb*{B}^2}{8\pi}\right)\vb*{I}-\frac{\vb*{B}\boldsymbol\otimes\vb*{B}}{4\pi}\biggr]=0,
\end{equation}
\begin{equation}
        \pdv{e}{t}+\boldsymbol\nabla\cdot\biggl[\left(e+p+\frac{\vb*{B}^2}{8\pi}\right)\vb*{v}-\frac{\vb*{B}(\vb*{v}\cdot\vb*{B})}{4\pi}-\kappa\boldsymbol\nabla T_\mathrm{gas}\biggr]=\mathcal{G},
\end{equation}
\begin{equation}
    \pdv{\vb*{B}}{t}-\boldsymbol\nabla\boldsymbol\cross(\vb*{v}\boldsymbol\cross\vb*{B})=0,
\end{equation}
and
\begin{equation}
    \pdv{n(\mathrm{A})}{t}+\nabla\cdot (n(\mathrm{A})\vb*{v})=R(\mathrm{A}).
\end{equation}
Here, $\rho$ is the gas mass density: 
\begin{equation}
    \rho=\mu_\mathrm{H} m_\mathrm{H}n_\mathrm{H},
\end{equation}
$\vb*{v}$ is the velocity, $p$ is the thermal pressure, $\vb*{B}$ is the magnetic field, $e$ is the total energy density:
\begin{equation}
    e=\frac{p}{(\gamma-1)}+\frac{1}{2}\rho |\vb*{v}|^2+\frac{1}{8\pi} |\vb*{B}|^2,
\end{equation}
$T_\mathrm{gas}$ is the gas temperature, $\kappa$ is the thermal conductivity, $\mathcal{G}$ is the net heating rate per unit volume, $n(\mathrm{A})$ is the number density of chemical species A, $R(\mathrm{A})$ is the net formation rate of species A, $\mu_\mathrm{H}=1.4$ is a factor that arises from the mass fraction of helium relative to hydrogen, $m_\mathrm{H}$ is the mass of a hydrogen atom, $n_\mathrm{H}$ is the number density of total hydrogen nuclei, and $\gamma=5/3$ is the heat capacity ratio.  We set the thermal conductivity to be that of hydrogen used in \citet{Parker_1953}. The equation of state is given as
\begin{equation}
    p=\frac{\rho}{\mu m_\mathrm{H}}k_\mathrm{B}T_\mathrm{gas},
\end{equation}
where $\mu$ is the mean molecular weight, and $k_\mathrm{B}$ is Boltzmann's constant. We used {\sc athena++} \citep{Stone_2020} to solve the above equations.

We adopted the chemical network and the heating and cooling functions used in \citet{Gong_2017, Gong_2023}. The chemical network includes 18 species (H, H$^+$, H$_2$, H$_2^+$, H$_3^+$, He, He$^+$, C, C$^+$, CH$_\mathrm{x}$, CO, HCO$^+$, O, O$^+$, OH$_\mathrm{x}$, Si, Si$^+$, and e$^-$) and 50 chemical reactions, where CH$_\mathrm{x}$ represents CH, CH$_2$, CH$^+$, CH$_2^+$, and CH$_3^+$, and OH$_\mathrm{x}$ does OH, H$_2$O, OH$^+$, H$_2$O$^+$, and H$_3$O$^+$. The reactions are all gas-phase reactions, except for H$_2$ formation on grain surfaces and grain-assisted recombination of ions. The elemental abundances of He, C, O, and Si are set to be the same as those used in \citet{Gong_2017} (see also Table \ref{tab:InitialAbundance}).

To evaluate the photodissociation/ionization rates and the photoelectric heating rates at each cell, we estimate $A_V$ and self-shielding factors of H$_2$, CO, and atomic C with column densities of total hydrogen nuclei and each molecule. We adopted the six-ray approximation to evaluate the attenuation of the interstellar radiation field \citep[e.g.][]{Glover_2010}. For example, the photodissociation rate of H$_2$, $D_\mathrm{pd}(\mathrm{H_2})$, at $(x,y,z)$ is calculated as
\begin{equation}\label{eq:H2shield}
    D_\mathrm{pd}(\mathrm{H_2})=\frac{1}{6}\sum_{\mathrm{i}=\pm x,\pm y,\pm z} \alpha(\mathrm{H_2})f_\mathrm{H_2}(N_\mathrm{H_2,i})\exp(-\beta(\mathrm{H_2})A_{V,\mathrm{i}}),
\end{equation}
where $\alpha(\mathrm{H_2})$ and $\beta(\mathrm{H_2})$ are the coefficients for the photodissociation rate, $f_{\rm H2}$ is the self-shielding factor of H$_2$ \citep{Draine_1996}, and $N_\mathrm{H_2,i}$ is the column density of H$_2$ calculated by integrating the number density from the $\mathrm{i}=\pm x, \pm y, \pm z$ boundaries. $A_{V,\mathrm{i}}$ is the visual extinction from each boundary that is calculated as
\begin{equation}\label{eq:Av}
    A_{V,\mathrm{i}}=\frac{N_\mathrm{H,i}}{1.87\times 10^{21}\,\mathrm{cm^{-2}}},
\end{equation}
where $N_\mathrm{H,i}$ is the column densities of total hydrogen nuclei, following \citet{Bohlin_1978} and \citet{Gong_2017}. We set the interstellar radiation field to be $G_0=1.0$ in the Habing unit \citep{Habing_1968}. The cosmic ray ionization rate of a hydrogen atom is set to be $\zeta_\mathrm{H}=2.0\times 10^{-16}\,\mathrm{s^{-1}}$ \citep[e.g.][]{Dalgarno_2006}.

\subsubsection{Calculation set-up}\label{sec:MHDsetup}

\HI gas is thought to have a two-phase structure: a mixture of warm neutral medium (WNM) and cold neutral medium (CNM). \citet{Koyama_2002} showed that this structure is naturally formed by interstellar shock waves. The timescale for ISM to be swept up by a shock wave from a supernova explosion is roughly 1 Myr \citep{1977ApJ...218..148M, Hennebelle_2019}, which is shorter than the typical life-time of molecular clouds, a few tens Myr. Therefore, it is natural to assume that molecular clouds should form from atomic clouds with a two-phase structure.

Here we describe how we set up the initial two-phase gas following the approach of \citet{Iwasaki_2019}. We set a uniform medium with a number density of total hydrogen nuclei of $n_\mathrm{H,0}=10\,\mathrm{cm^{-3}}$ and temperature of 400 K in a cubic box with a side length of $L=10\,\mathrm{pc}$. The initial chemical condition is set to be completely atomic. To force thermal instability, we put subsonic Kolmogorov turbulence with the velocity dispersion of 17\% of the initial sound speed, which value is similar to that of \citet{Iwasaki_2019}. The strength of the initial magnetic field is $5\,\mathrm{\mu G}$. The angle between the initial magnetic field and the $x$-axis is denoted by $\theta$. The choice of $\theta$ is described later. We adopt the optically thin approximation; shielding of the radiation field by dust and molecules is excluded in this calculation for setting the initial condition, e.g., the visual extinction and the molecular column densities are set to be zero when calculating the heating, cooling, and photodissociation rates. To save computational time, the calculation region is divided into $128\times 128\times 128$ cells. A high spatial resolution is not required here because the purpose of this calculation is solely to obtain the steady-state structure of the atomic gas. We continue the calculation until 10 Myr, when the structure of the medium is in a steady-state.

\begin{figure}
	\includegraphics[width=\columnwidth]{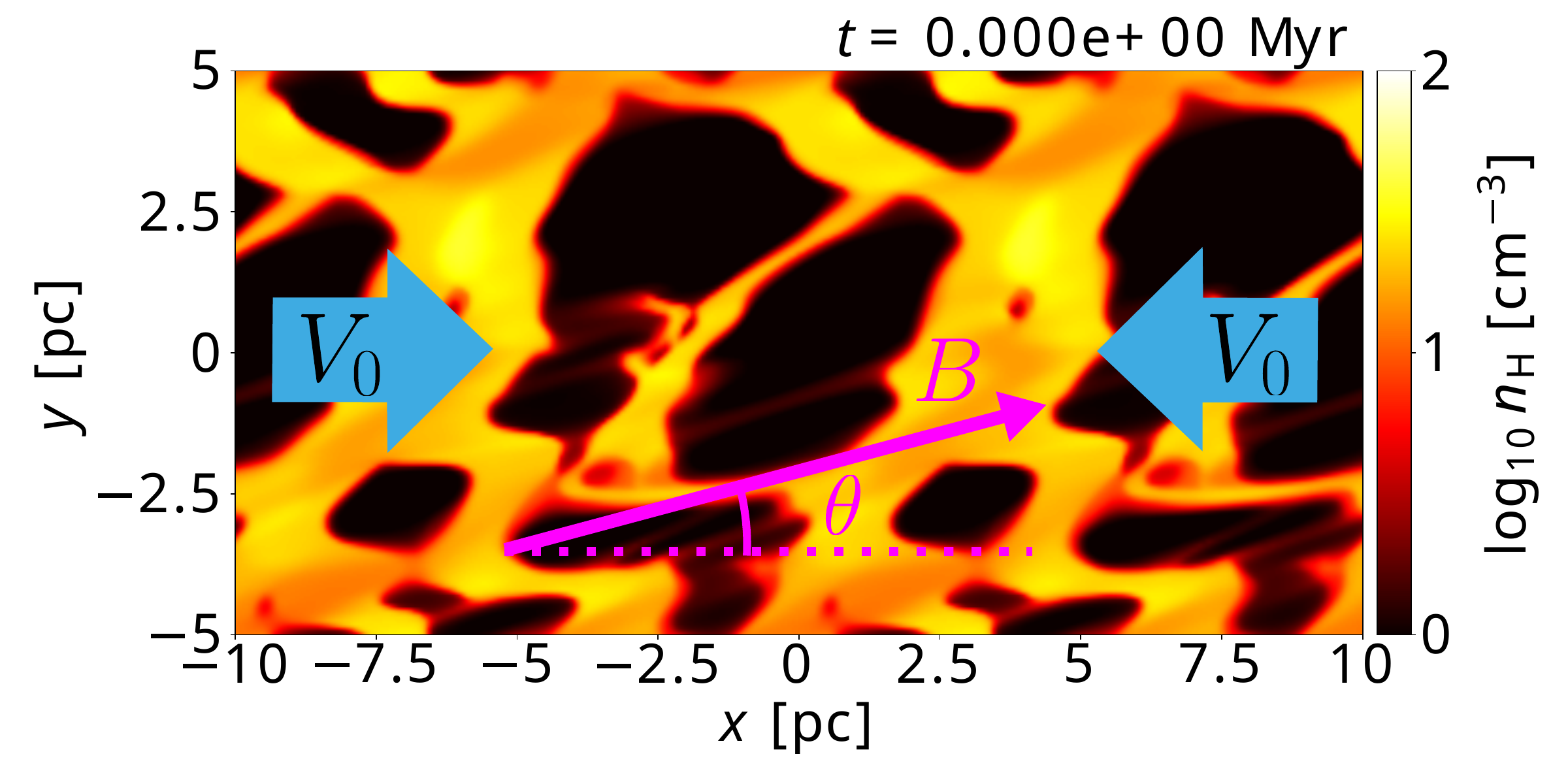}
    \caption{The initial condition of the converging flow simulation. The color shows the number density of hydrogen nuclei on the $z=0$ plane.}
    \label{fig:initial_condition}
\end{figure}

For the initial condition of the colliding flow, the calculation region described above is divided into $256 \times 256 \times 256$ by interpolation, and then doubled along the $x$-axis as shown in Figure \ref{fig:initial_condition}. The calculation domain is 20 pc $\times$ 10 pc $\times$ 10 pc, and the total number of cells is $512 \times 256 \times 256$. The spatial resolution meets the requirement for the convergence of H$_2$ and CO abundance \citep[e.g.][]{Joshi_2019}. The initial velocity field is set as follows to let the medium collide at the $x=0$ plane;
\begin{equation}
    v_x=-V_0\tanh\left({\frac{x}{0.5\,\mathrm{pc}}}\right),
\end{equation}
where $V_0$ is set to be 10\,km\,s$^{-1}$. The boundary condition at the $x=\pm 10\,\mathrm{pc}$ plane is set so that the initial medium continuously inflows with velocity $V_0$. We set periodic boundary conditions for the $y$ and $z$ directions.

As shown by \citet{Iwasaki_2019}, the initial angle of the magnetic field $\theta$ against the $x$-axis determines the evolution of compression layers. They found a critical angle for the efficient compression of atomic gas to form molecular clouds. When the angle is larger than the critical one, the magnetic pressure prevents the compression. And if the angle is too small, strong turbulence due to the non-linear thin-shell instability makes the volume filling factor of the dense region tiny. In the present work, we choose $\theta$ to be the critical value, $18^\circ$, so that we focus on the efficient molecular cloud formation condition.

We put tracer particles in the calculation domain and analyze the time evolution of physical quantities of each fluid element. We distribute particles so that they continuously get into the compression layer at almost a constant rate. At the first timestep of the calculation, we distribute $64\times 64 \times 48$ particles randomly in the calculation domain. Then we injected one tracer particle from the calculation boundary at $x= 10\,\mathrm{pc}$ and $x=- 10\,\mathrm{pc}$ every $1.2\times 10^{-5}$ Myr so that the total number of tracer particles at 5\,Myr is about $10^6$. The initial location of the injected particles on $yz$ plane is randomly determined. Using the location of tracer particles and the physical quantities of neighboring calculation cells, we calculated the density, temperature, and column densities of total hydrogen nuclei, H$_2$, CO, and C from the boundaries by linear interpolation. The interpolation is conducted every $1.0\times 10^{-4}$ Myr, which is sufficiently small compared with the timescale for a typical sound wave to cross a calculation cell in the compression layer. We also calculated the molecular abundances at the position of the tracer particle by interpolation of their logarithm between the neighbouring cells. To solve the trajectories of tracer particles, we used the Lagrangian particle module of {\sc athena++} (C.-C. Yang et. al. in preparation).

The simulation continues until 5 Myr, when the typical visual extinction reaches 1 mag and the compression layer becomes steady-state. The total number of tracer particles is 1029938. We select one dense clump and its surrounding region (see Section \ref{sec:MHDresult}) and pick up 20,000 particles within it for the post-processing chemical network calculation that is explained in the following section.

\subsection{Chemical network calculation}\label{sec:Chem}

\subsubsection{Chemical network}\label{sec:ChemSetup}

To simulate the chemical evolution along the trajectory of tracer particles with a large chemical network, we used {\sc rokko}, a chemical network calculation code described in \citet{Furuya_2015}. This code calculates the time evolution of chemical abundances using the rate equation method with a three-phase model considering the gas phase, grain surface, and inert bulk mantle \citep{Hasegawa_1993b}. 

A set of chemical reactions and their rates are similar to those used in \citet{Komichi_2024}, which is based on a kinetic database for astrochemistry (KIDA; \citet{Wakelam_2012}). We additionally include recombination reactions of ions with polycyclic aromatic hydrocarbons (PAHs) following \citet{Wakelam_2008} and \citet{1987ApJ...320..803D}. The total number of chemical species and reactions are 953 and 12029, respectively. We assumed the ratio of diffusion barrier to desorption energy to be 0.4. The chemical desorption probability per reaction is set to be 1 per cent \citep{Garrod_2007}. Following \citet{Furuya_2015}, we set the photodesorption yield for CO as a function of the surface coverage of CO ice and the yield for others to be $10^{-3}$.

There are three updates in the calculation setup of the post-process chemical network calculations from that of \citet{Komichi_2024}. Firstly, we modify the size distribution of dust grains. If we simply adopt a grain radius of $1\times 10^{-5}$ cm, which is a commonly assumed value, the total surface area of dust grains would be underestimated. While the grain size is set to be a single value in our chemical network calculation for simplicity, we set the grain radius so that the total surface area of dust grains is consistent with that of \citet{Mathis_1977}. We adopt the size range of $0.005\,\mathrm{\mu m}$ to $0.25\,\mathrm{\mu m}$ and the power-law size distribution with an exponent of $-3.5$. The effective grain radius corresponding to the total cross section is $3.54\times 10^{-6}$ cm, which is about three times smaller than $1\times 10^{-5}$ cm. The dust-to-gas mass ratio is set to 0.01.

Secondly, we reduced the number of the two-body reactions on grain surfaces. In the present work, we are mainly interested in gas-phase molecules and simple molecular ices on dust grains, rather than organic molecules larger than methanol. To save computational time, we reduce the number of two-body reactions from 591 to 244 so that most of them are hydrogenation reactions except for reactions relevant to the formation of H$_2$O (H$_2$ + OH $\rightarrow$ H$_2$O + H), CO (C + O $\rightarrow$ CO and O + HCO $\rightarrow$ CO + OH), and CO$_2$ (OH + CO $\rightarrow$ CO$_2$ + H and O + HCO $\rightarrow$ CO$_2$ + H). The formation rates of molecular ice by the two-body reactions are calculated using the diffusivity of each chemical species based on the Langmuir-Hinshelwood mechanism. We conducted a chemical network calculation using this reduced surface chemical network and compared the result with that of the fiducial model of \citet{Komichi_2024} to confirm that this reduction does not have a large effect on abundances of major molecular ices such as H$_2$O and CO$_2$.

Thirdly, we modified the H$_2$ formation rate via the recombination of two atomic H on grains. In the simple chemical network model implemented in our MHD simulation, the H$_2$ formation rate is given as $R_\mathrm{H_2}n_\mathrm{H}n(\mathrm{H)}$, where $n(\mathrm{H})$ is the number density of atomic hydrogen, and $R_\mathrm{H_2}=3\times 10^{-17}\,\mathrm{cm^{3}\,s^{-1}}$ is the rate coefficient of H$_2$ formation, which is a typical value to reproduce the observations of diffuse clouds derived by e.g. \citet{Jura_1975} \citep[see also][]{Kaufman_1999,Wolfire_2008}. 
It is assumed that no molecular ice forms on the surface, i.e., all adsorbed H atoms are consumed for H$_2$ formation, which is reasonable for diffuse clouds. As the fluid elements migrate to dense cold regions, however, molecular ices start to form, which competes with H$_2$ formation. In addition, the surface energy potential of the icy surfaces is different from that of the bare grains. Therefore, we separately evaluate the H$_2$ formation rate on the bare grain and the ice surface. Assuming that adsorbed H atoms on bare grain are immediately converted to H$_2$, we set the H$_2$ formation rate on the bare grain as follows:
\begin{equation}
    R_\mathrm{H_2}^\mathrm{bare}=\max(0, R_\mathrm{H_2}-S\sigma_\mathrm{gr}v_\mathrm{th}n_\mathrm{gr}\theta_\mathrm{ice}/n_\mathrm{H}), 
\end{equation}
where $S$ is the sticking coefficient of a hydrogen atom \citep{Hollenbach_1979}, $\sigma_\mathrm{gr}$ is the cross section of a dust grain, $v_\mathrm{th}$ is the thermal velocity, $n_\mathrm{gr}$ is the number density of dust grains, and $\theta_\mathrm{ice}$ is the surface coverage of molecular ice. As molecular ices are formed on the grain surface (i.e., $\theta_\mathrm{ice}$ increases), the rate of H$_2$ formation on the bare grain decreases. H$_2$ formation rate on the ice surface is evaluated by solving the competition between the formation of H$_2$ and other molecules based on the Langmuir-Hinshelwood mechanism.

\begin{table}
	\centering
	\caption{Initial abundances used in the dynamical model. }
	\label{tab:InitialAbundance}
	\begin{tabular}{lc} 
		\hline
        H       & $1.00$     \\
        He      & $1.00(-1)$ \\
        N       & $7.60(-5)$ \\
        O       & $3.20(-4)$ \\
        C$^+$   & $1.60(-4)$ \\
        S$^+$   & $1.50(-5)$ \\
        Si$^+$  & $1.70(-6)$ \\
        Fe$^+$  & $2.00(-7)$ \\
        Na$^+$  & $2.00(-7)$ \\
        Mg$^+$  & $2.40(-7)$ \\
        Cl$^+$  & $1.80(-7)$ \\
        P$^+$   & $1.17(-7)$ \\
        F$^+$   & $1.80(-8)$ \\
		\hline
	\end{tabular}
\end{table}

\subsubsection{Model description}\label{sec:ChemModel}

From our MHD simulations, we obtain gas density, gas temperature, visual extinction (equation (\ref{eq:Av})), and self-shielding factors of H$_2$, CO, and C (e.g., equation (\ref{eq:H2shield})) of each tracer particle as functions of time. The dust temperature is given as a function of visual extinction with the equation (8) of \citet{Hocuk_2017}. 
Following \citet{Glover_2010} \citep[see also e.g.][]{Offner_2013}, we adopted the effective visual extinction as follows:
\begin{equation}
    \langle A_V\rangle=-\frac{1}{\gamma}\ln \left[\frac{1}{6}\sum_{\mathrm{i}=\pm x,\pm y,\pm z}\exp(-\gamma A_{V,\mathrm{i}}) \right],
\end{equation}
where $\gamma=1.87$ is the coefficient for the dust extinction \citep{Gong_2017}.
We conduct chemical network calculations using the temporal variation of physical quantities. The initial abundance is listed in Table \ref{tab:InitialAbundance}. The elemental abundances of He, C, O, and Si are the same as those used in the MHD simulations. Those of other elements are taken from EA2 model in \citet{Wakelam_2008}. We refer to this model as a "dynamical model" hereinafter. 

To examine the effects of the physical history of particles on their chemical evolution, we calculate two additional models: "static model" and "H$_2$-fixed model". In these two models, the parameters for each tracer particle, i.e., gas density, gas temperature, dust temperature, visual extinction, and self-shielding factors for H$_2$, CO, and C, are set to be the values at the final timestep of the dynamical models and are kept constant. We set the initial abundance as in Table \ref{tab:InitialAbundance}, except for hydrogen, for which initial atomic and molecular abundances are taken from the final timestep of the dynamical model. The calculations are conducted for 5\,Myr for all particles.
While the chemistry of molecular clouds is non-equilibrium in general, it reaches a steady-state in a relatively short timescale at low $A_{\rm V}$, where photodissociation dominates in the chemical network \citep[e.g.][]{Yamamoto_2017}. We thus naively expect that the results of the static model and dynamical model may show reasonable agreement at 5 Myr for tracer particles in low $A_{\rm V}$ regions. 

We note, however, that H$_2$ abundance needs special attention. H$_2$ can be self-shielded from UV radiation, while its formation timescale is relatively long compared to the dynamical timescale, especially at low gas densities. H$_2$ abundance would not reach a steady-state value in a short timescale even at low $A_{\rm V}$. Indeed, the previous works \citep[e.g.][]{Valdivia_2016} show that H$_2$ formation is affected by the physical evolution of fluid elements. Furthermore, H$_2$ abundance could affect the whole chemical network, since the formation of interstellar molecules other than H$_2$ is initiated by two-body reactions of atoms with H$_2$ or H$_3^+$. Therefore, the non-equilibrium (non-steady-state) H$_2$ abundance, in other words, the temporal and local H$_2$ abundance, can be essential for the abundances of other molecules. In the static model, the initial H$_2$ abundance is set to the value at the final timestep of the dynamical model, but the H$_2$ abundance varies with time and asymptotically approaches its steady-state value at the local physical condition. In order to investigate the effect of H$_2$ abundance on other molecules, we kept the abundance of H$_2$ constant (i.e., the same as that of the final timestep of the dynamical model) in the H$_2$-fixed model by modifying the reaction rates for H$_2$ as follows. We ignore the H$_2$ destruction by photodissociation. We set the conversion rate of gas-phase H atoms to H$_2$ equal to the destruction rate of H$_2$ by cosmic rays, which is the major destruction path if the photodissociation is neglected. We also modify the products of the chemical desorption of two-body reaction H(s) + H(s) and the photodesorption of H$_2$(s) from H$_2$(g) to 2H(g), where (g) and (s) denote gas-phase and surface species, respectively. As shown in Section \ref{sec:dynamics_static}, this simple treatment can keep the H$_2$ abundance unchanged from the final abundance of the dynamical model. 



\section{Results}\label{sec:Results}

\subsection{Overview of the MHD simulations}\label{sec:MHDresult}

\begin{figure}
	\includegraphics[width=\columnwidth]{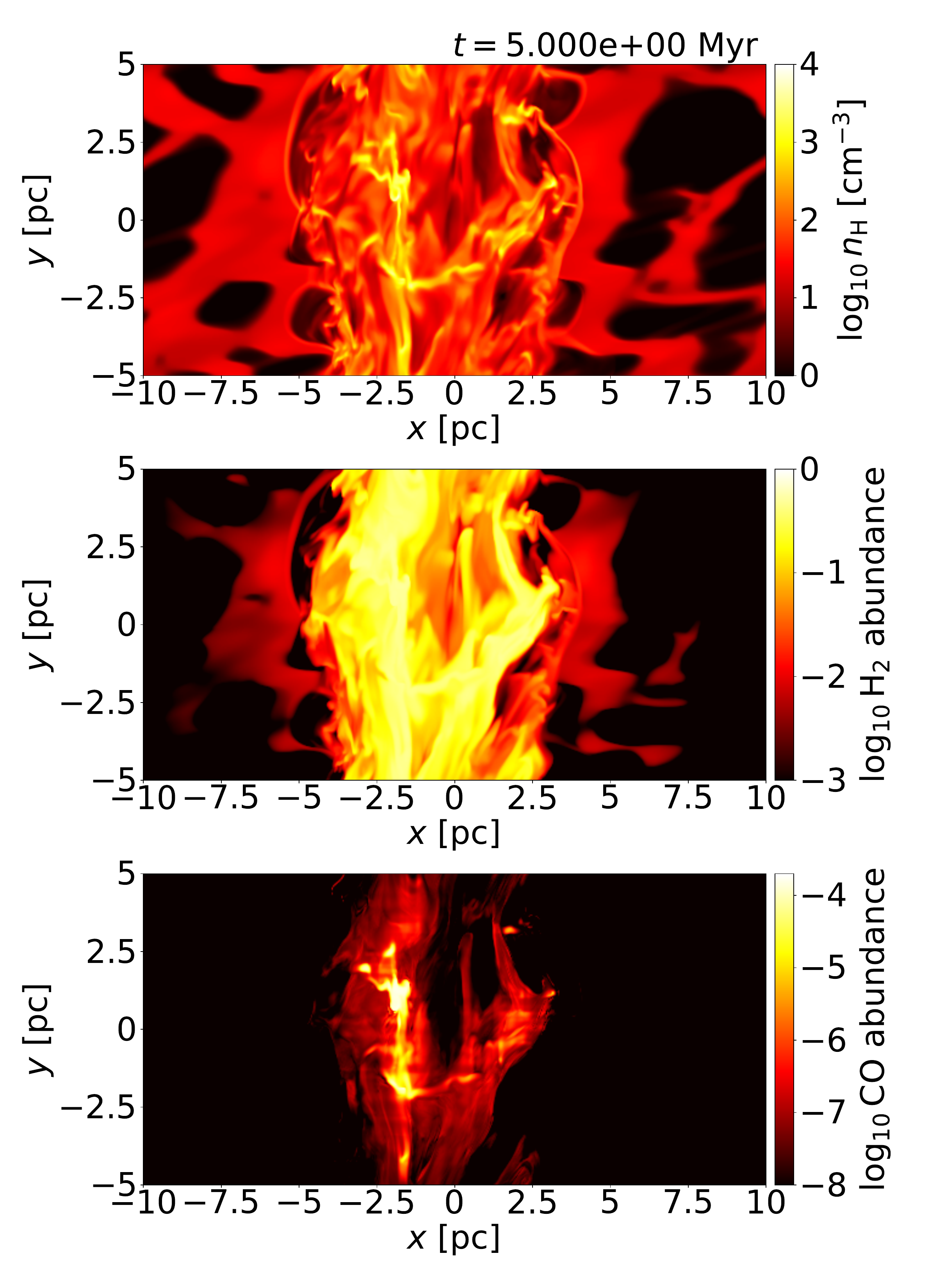}
    \caption{Slices of gas density, H$_2$ abundance, and CO abundance in the $z=0$ plane at 5 Myr.}
    \label{fig:slice_summary_5Myr}
\end{figure}

In this section, we show the overall results of the MHD simulations. Figure \ref{fig:slice_summary_5Myr} shows the number density, H$_2$ abundance, and CO abundance on the $z=0$ plane at 5 Myr. The compression layer, where the gas density is higher than in the pre-shock gas, is formed in the central region of $|x|< 5$ pc. H$_2$ is formed throughout the compression layer. On the other hand, CO abundance is saturated ($\sim 10^{-4}$) only in limited regions with high gas density ($n_{\rm H}> 10^3$ cm$^{-3}$). This is because the CO formation requires shielding from the external UV radiation field by dust grains \citep[e.g.][]{2004ApJ...612..921B}. The typical column density of total hydrogen nuclei from the boundary to the center of the compression layer at $5\,\mathrm{Myr}$ is about $n_\mathrm{H,0}V_0\times 5\,\mathrm{Myr}\simeq 1.5\times 10^{21}\,\mathrm{cm^{-3}}$, i.e., $A_V\simeq 0.8\,\mathrm{mag}$. This value corresponds to the threshold visual extinction for CO formation \citep{2004ApJ...612..921B}. Therefore, a significant fraction of the early stage of molecular clouds can be a CO-poor molecular gas, where the \HI-to-H$_2$ transition is completed but CO is not well formed. In the present work, molecular clouds in which the CO abundance relative to H$_2$ is well below $10^{-4}$ and the abundances of H$_2$ and CO show no correlation are referred to as 'CO-poor' clouds. Our definition of CO-poor clouds does not necessarily coincide with that of CO-dark clouds (i.e. molecular clouds with weak CO emission; e.g. \citet{vanDishoeck_1992}), as we do not calculate line intensities.

\begin{figure}
	\includegraphics[width=\columnwidth]{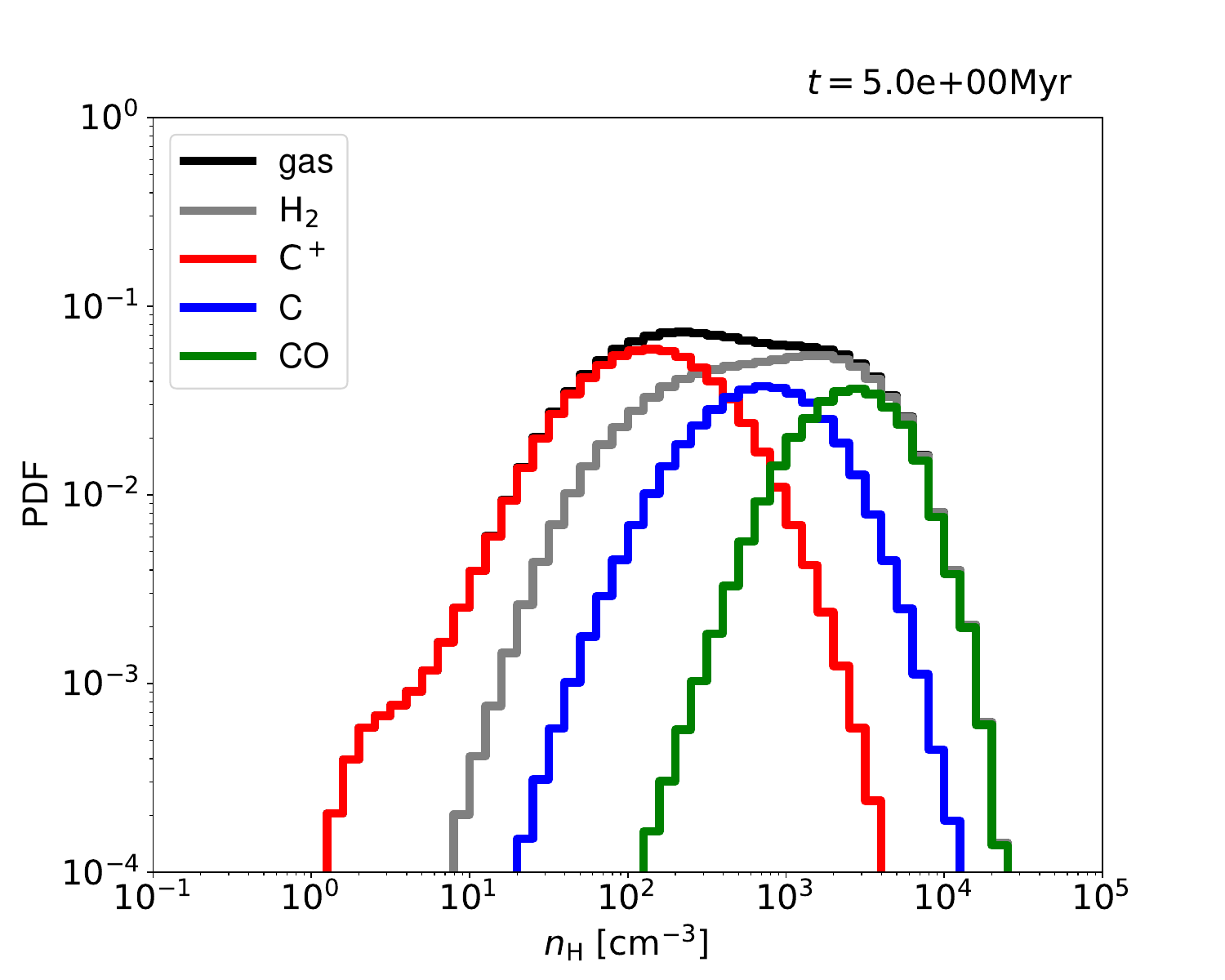}
    \caption{Mass-weighted probability distribution function of the gas density, H$_2$, and major carbon carriers at 5 Myr.}
    \label{fig:chemPDF}
\end{figure}

Figure \ref{fig:chemPDF} shows the mass-weighted probability distribution function (PDF) of gas density, H$_2$, and major carbon carriers in the compression layer at $5\,\mathrm{Myr}$. The definitions of PDFs are the same as equations (14), (17), and (18) of \citet{Iwasaki_2019}; 
\begin{equation}
    \mathrm{PDF_{gas}}(s_\mathrm{i})=\frac{1}{M_\mathrm{tot}\Delta s}\int_{|s-s_\mathrm{i}|<\Delta s/2}\rho \,\mathrm{d}^3x,
\end{equation}
\begin{equation}
    \mathrm{PDF_{H_2}}(s_\mathrm{i})=\frac{1}{M_\mathrm{tot}\Delta s}\int_{|s-s_\mathrm{i}|<\Delta s/2}2\mu_\mathrm{H} n(\mathrm{H_2})m_\mathrm{H} \,\mathrm{d}^3x,
\end{equation}
and,
\begin{equation}
    \mathrm{PDF_{X_C}}(s_\mathrm{i})=\frac{1}{A_\mathrm{C}M_\mathrm{tot}\Delta s}\int_{|s-s_\mathrm{i}|<\Delta s/2}\mu_\mathrm{H} n(\mathrm{X_C})m_\mathrm{H} \,\mathrm{d}^3x,
\end{equation}
where $s=\log_{10}n_\mathrm{H}$ is divided into uniform bins. $\mathrm{X_C}$ denotes C$^+$, C, or CO, $\Delta s$ is the width of the bins, and $A_\mathrm{C}$ is the elemental abundance of carbon. $M_\mathrm{tot}=2\mu_\mathrm{H} m_\mathrm{H}n_\mathrm{H,0}V_0L^2t$ is the total mass of the compression layer at time $t$. More than 10\% of hydrogen already exists as H$_2$ at $n_\mathrm{H}>10\,\mathrm{cm^{-3}}$, whereas 10\% of carbon exists as CO at $n_\mathrm{H}> 3\times 10^2\,\mathrm{cm^{-3}}$. H$_2$ and CO abundances are correlated only in higher-density regions of the molecular clouds in the formation stage. 
This again indicates that CO formation remains incomplete in many regions of a molecular cloud during its formation phase.

\begin{figure*}
	\includegraphics[width=2.0\columnwidth]{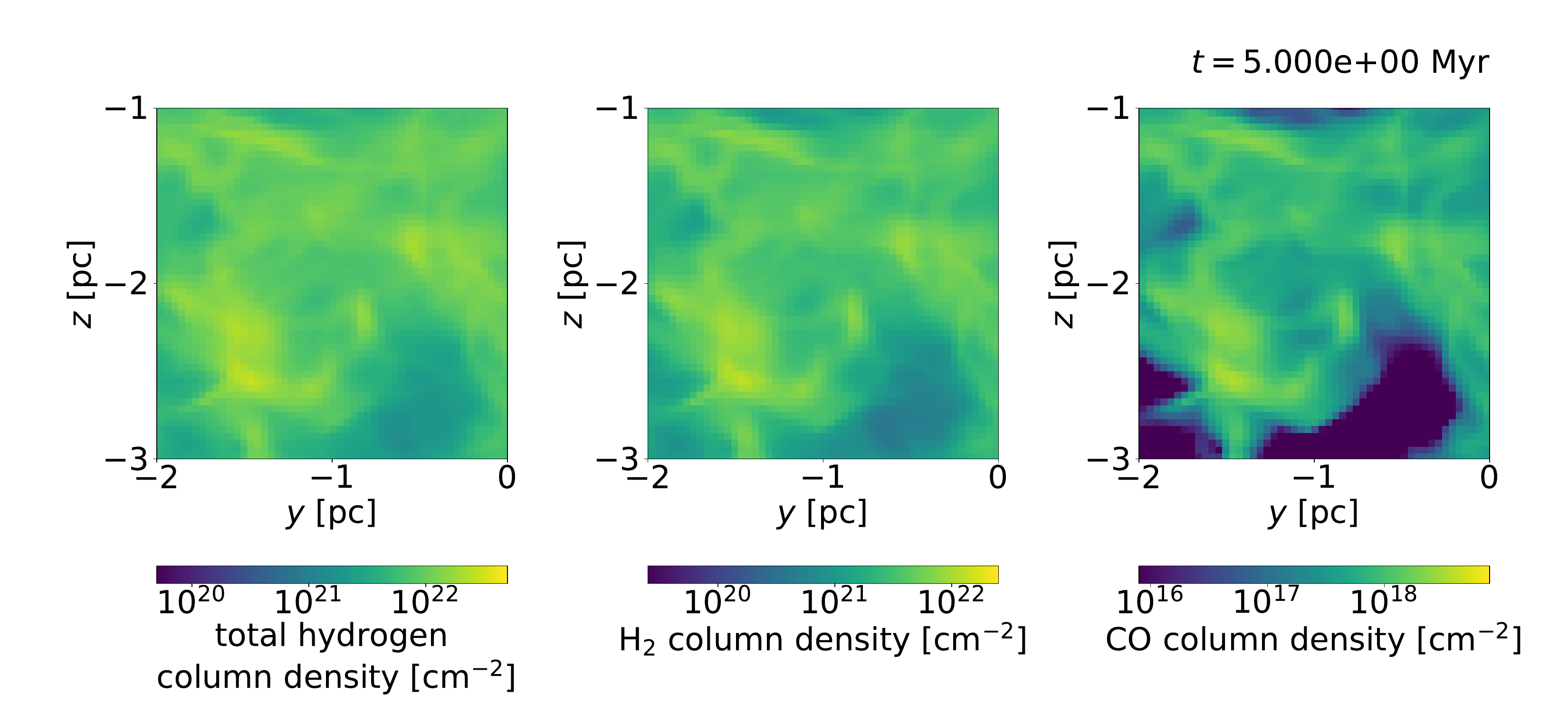}
    \caption{The total hydrogen column density, H$_2$ column density, and CO column density along $x$-axis at 5 Myr.}
    \label{fig:clump_column}
\end{figure*}

For the post process calculation of large chemical reaction network, we focus on one dense clump and its surrounding region. We selected the region with $x\in [-10\,\mathrm{pc}, 10\,\mathrm{pc}]$, $y\in [-2\,\mathrm{pc}, 0\,\mathrm{pc}]$, and $z\in [-3\,\mathrm{pc}, -1\,\mathrm{pc}]$. Figure \ref{fig:clump_column} displays the column density of the total hydrogen nuclei, H$_2$, and CO on the $yz$ plane of the selected region. The densest region exists around $y\sim -1.5\,\mathrm{pc}$ and $z\sim -2.5\,\mathrm{pc}$ on the $yz$ plane, where both H$_2$ and CO abundances are almost saturated. On the other hand, there is a CO-poor region around $y\sim -0.5\,\mathrm{pc}$ and $z\sim -2.5\,\mathrm{pc}$. 

\begin{figure}
	\includegraphics[width=\columnwidth]{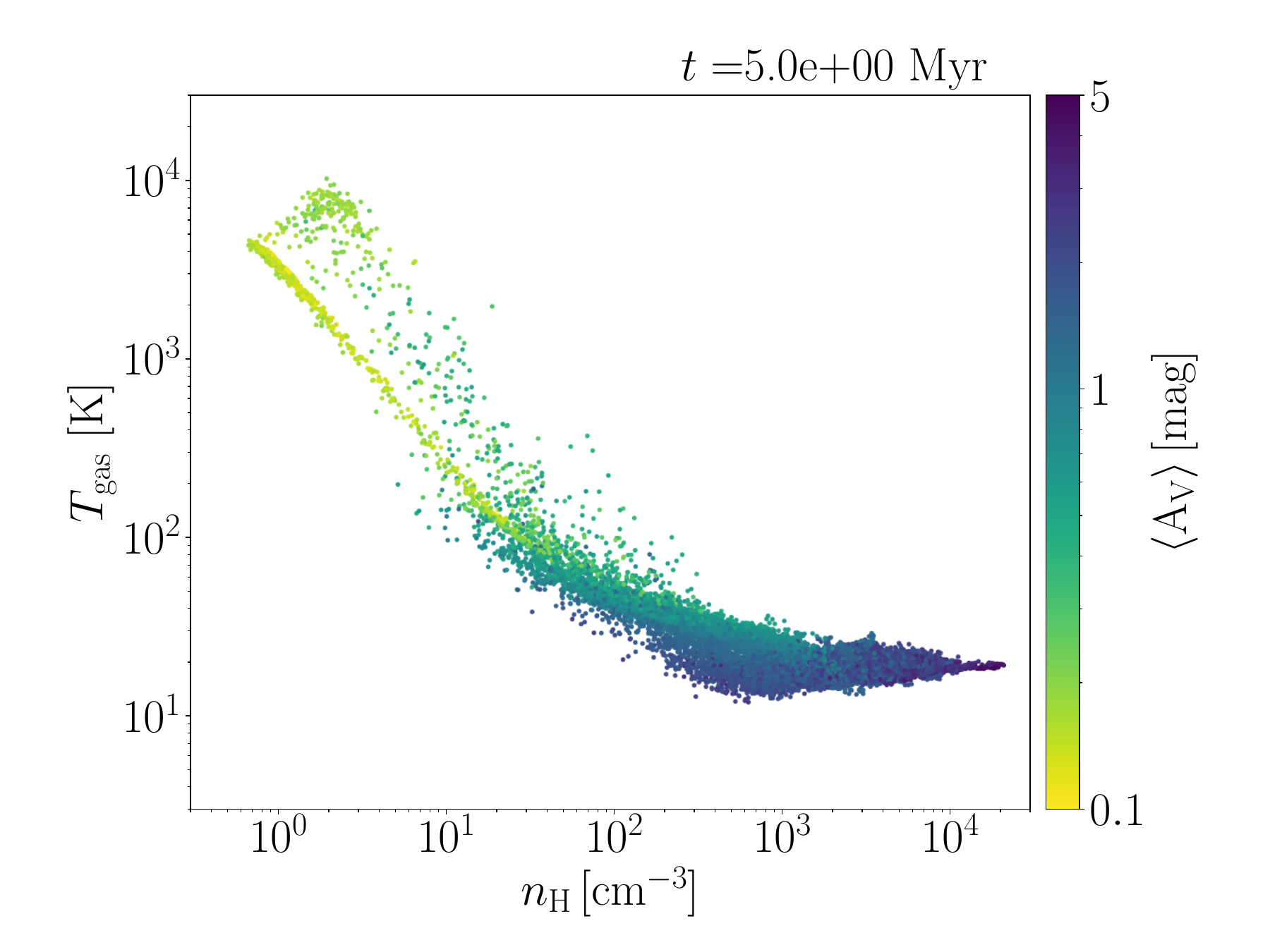}
    \caption{The gas density, temperature, and $A_V$ of each tracer particles at 5 Myr.}
    \label{fig:phys_tracer}
\end{figure}

We pick up 20,000 tracer particles in this region. Figure \ref{fig:phys_tracer} shows the gas number density, gas temperature, and effective visual extinction $\langle A_V\rangle$ of the particles at the final timestep. The effective visual extinction ranges from 0.1 mag (yellow) to 5 mag (purple). The yellow dots, almost aligned in a straight line at $n_\mathrm{H}\sim 1-30\,\mathrm{cm^{-3}}$, show the parameters of the tracer particles in the preshock region. The gas temperature of the dense region (i.e., $n_\mathrm{H}\gtrsim 10^3\,\mathrm{cm^{-3}}$) is about $20\,\mathrm{K}$, which is slightly higher than the typical gas temperature of molecular clouds, $10\,\mathrm{K}$. This is because of heating by cosmic rays with the assumed high ionization rate of $2\times 10^{-16}\,\mathrm{s^{-1}}$ \citep[e.g.][]{Clark_2019}. At a given density, the gas temperature increases as the effective visual extinction decreases due to photoelectric heating.

\subsection{Molecular hydrogen}\label{sec:H2}
We find good agreement of molecular abundances for the major species that are common to both the small chemical network implemented in the MHD calculation and the large network calculation of the dynamical model (Appendix \ref{app:otf_ppc}).
Hereafter, we show the results of post-process chemical network calculations. 

\begin{figure}
	\includegraphics[width=\columnwidth]{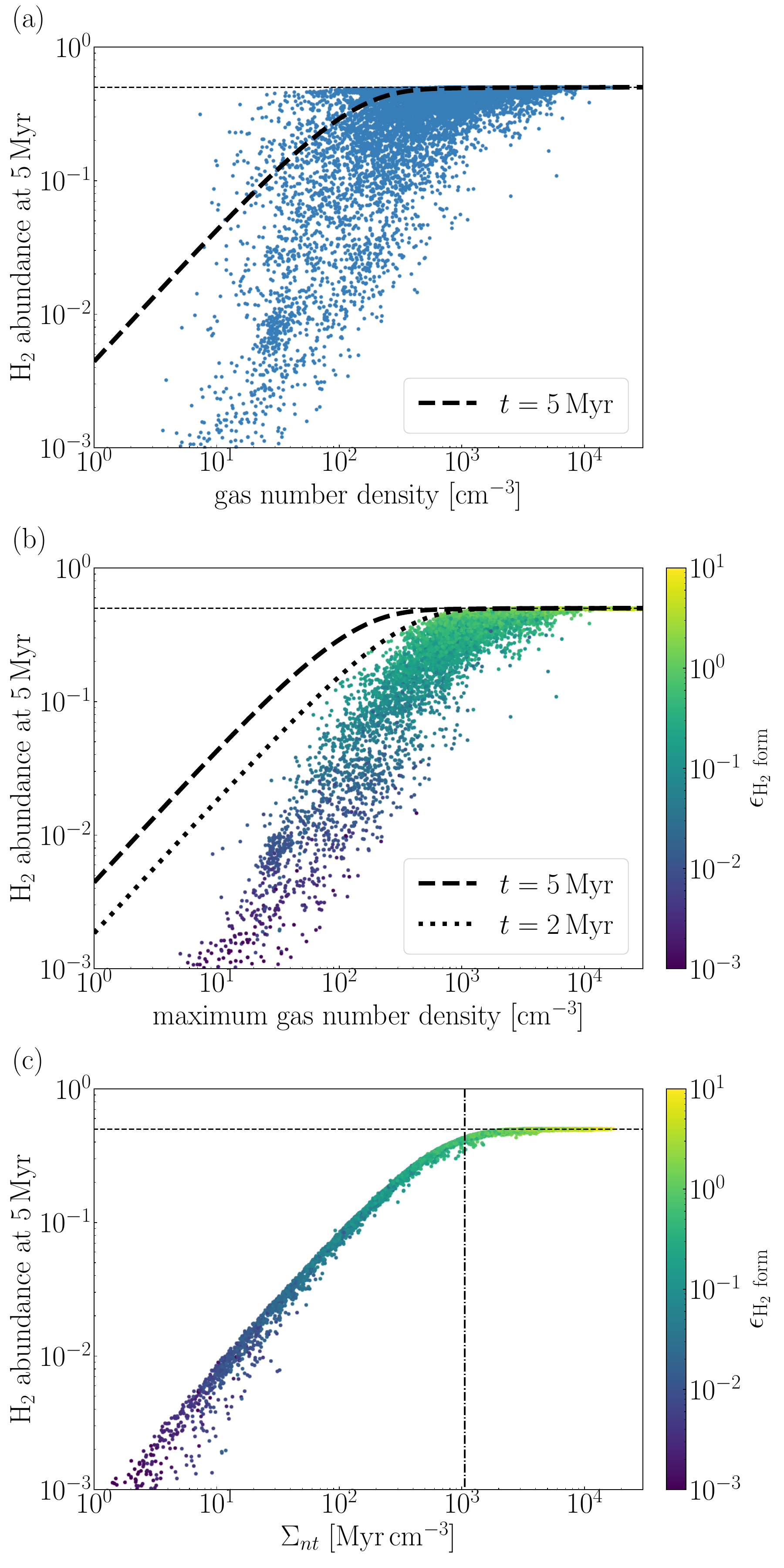}
    \caption{(a) gas number density and H$_2$ abundance of each tracer particle at 5\,Myr. The dashed line shows the analytic solution of H$_2$ abundance (equation (\ref{eq:H2analytic})) at 5\,Myr. (b) The maximum gas number density that each tracer particle experiences ($n_\mathrm{H,\,max}$) and the H$_2$ abundance at 5\,Myr. The color scale shows the efficiency of H$_2$ formation, $\epsilon_\mathrm{H_2\ form}$, defined by equation (\ref{eq:H2formeff}). The dotted line shows the analytic solution of H$_2$ abundance at 2\,Myr. (c) The time integration of gas number density ($\Sigma_{nt}$) and the H$_2$ abundance of each tracer particle.}
    \label{fig:H2_summary}
\end{figure}

We first focus on H$_2$, which is the most fundamental molecule of molecular clouds. Figure \ref{fig:H2_summary}(a) shows the gas number density and H$_2$ abundance of each tracer particle at 5\,Myr. The dotted line shows the analytic solution of H$_2$ abundance at 5\,Myr. When H$_2$ is shielded from UV radiation, it is mainly destroyed by cosmic rays, H$_2$ + cosmic rays $\rightarrow$ H$_2^+$ + e$^-$, and the two-body reaction with H$_2^+$, H$_2$ + H$_2^+$ $\rightarrow$ H$_3^+$ + H. If the physical condition is constant and all hydrogen is initially atomic, the abundance of H$_2$ at time $t$, $a_\mathrm{H_2}(t)$, is given as follows:
\begin{equation}\label{eq:H2analytic}
    a_\mathrm{H_2}(t)=\frac{R_\mathrm{H_2}n_\mathrm{H}}{2(R_\mathrm{H_2}n_\mathrm{H}+\zeta_\mathrm{H_2})}\left[1-\exp\left\{-2(R_\mathrm{H_2}n_\mathrm{H}+\zeta_\mathrm{H_2})t\right\}\right],
\end{equation}
where $\zeta_\mathrm{H_2}$ is the cosmic ray ionization rate of H$_2$. Here, we assumed that H$_2^+$ abundance is determined by the balance between cosmic ray ionization of H$_2$ and the two-body reaction of H$_2^+$ with H$_2$. While the analytic solution of equation (\ref{eq:H2analytic})  gives the H$_2$ abundance as a function of gas density and time, the tracer particles in panel (a) exhibit a large scatter over an order of magnitude for a given gas density at $n_\mathrm{H}< 10^3\,\mathrm{cm^{-3}}$, with some particles showing higher H$_2$ abundances than the analytic solution. These differences are natural, because the analytic solution is obtained by assuming a constant gas density over 5 Myrs, while the gas density varies along the trajectory for tracer particles. In other words, H$_2$ abundance is not determined by the physical conditions of the final timestep but by the evolution of the gas density of each fluid element. 

Here, we define two values that characterize the history of the gas density. One is the maximum gas number density, $n_\mathrm{H,\,max}$, that each tracer particle experiences during the total calculation time. The other is the total time that each particle spends at gas densities higher than half of its maximum gas density, $t(n_\mathrm{H}>0.5n_\mathrm{H,\,max})$. This value represents the timescale during which each fluid element experiences dense conditions. Then the efficiency of H$_2$ formation during these high-density phases, $\epsilon_\mathrm{H_2\ form}$ , is described as 
\begin{equation}\label{eq:H2formeff}
    \epsilon_\mathrm{H_2\ form}=\frac{t(n_\mathrm{H}>0.5n_\mathrm{H,\,max})}{t_\mathrm{form,\,max}},
\end{equation}
where $t_\mathrm{form,\,max}=(R_\mathrm{H_2}n_\mathrm{H,\,max})^{-1}$ is the H$_2$ formation timescale at the maximum gas number density. Panel (b) of Figure \ref{fig:H2_summary} shows the maximum gas number density and H$_2$ abundance at 5\,Myr, and the color scale shows $\epsilon_\mathrm{H_2\ form}$. Compared to Panel (a), H$_2$ abundance correlates with the maximum gas number density. At the fixed maximum gas number density around $10\,\mathrm{cm^{-3}}<n_\mathrm{H,\,max}<10^3\,\mathrm{cm^{-3}}$, H$_2$ abundance is high when $\epsilon_\mathrm{H_2\ form}$ is large. As $\epsilon_\mathrm{H_2\ form}>1$ means that tracer particles stay in dense regions for a longer time than the H$_2$ formation timescale, H$_2$ abundances of particles with $\epsilon_\mathrm{H_2\ form}>1$ at $n_\mathrm{H,\,max}>10^3\,\mathrm{cm^{-3}}$ are closed to be saturated.  The dotted line in Panel (b) shows the analytic solution at 2\,Myr. We chose this time because 2\,Myr is the upper limit of $t(n_\mathrm{H}>0.5n_\mathrm{H,\,max})$ for almost all tracer particles; only 0.1\% of all tracer particles shows $t(n_\mathrm{H}>0.5n_\mathrm{H,\,max})>2\,\mathrm{Myr}$. Compared to the dashed line, the analytic solution at 5\,Myr, the dotted line gives more exact upper limit of H$_2$ abundances.  Therefore, as expected from equation (\ref{eq:H2analytic}), H$_2$ is mainly formed in dense regions. If photodissociation of H$_2$ is inefficient due to H$_2$ self-shielding or dust shielding, the destruction of H$_2$ is inefficient even in diffuse regions. Therefore, H$_2$ abundance is kept high around $n_\mathrm{H}<10^2\,\mathrm{cm^{-3}}$ at 5\,Myr (Panel (a) in Figure \ref{fig:H2_summary}). 
In addition, H$_2$ abundance is determined by the collision between H atoms and dust grains when the photodissociation is neglected. Thus, H$_2$ abundance should depend on the time integration of gas density, $\Sigma_{nt}$;
\begin{equation}
    \Sigma_{nt}=\sum_\mathrm{j}\,n_\mathrm{H}(\mathrm{j})\Delta t(\mathrm{j}),
\end{equation}
where $\mathrm{j}$ is a timestep and $\Delta t(\mathrm{j})$ is a time interval of each timestep $\mathrm{j}$. Panel (c) shows $\Sigma_{nt}$ and H$_2$ abundance of tracer particles, and the vertical dot-dashed line shows the reciprocal of the H$_2$ formation rate coefficient, $R_\mathrm{H_2}^{-1}\simeq 1\times 10^3\,\mathrm{Myr\,cm^{-3}}$. H$_2$ abundance linearly increases with $\Sigma_{nt}$ at $\Sigma_{nt}<R_\mathrm{H_2}^{-1}$, while it saturates at $\Sigma_{nt}>R_\mathrm{H_2}^{-1}$, which means that fluid elements stay in dense regions for a longer time than the H$_2$ formation timescale.
In summary, H$_2$ abundance is determined by the density history of each fluid element, which directly supports the statements of the previous works; H$_2$ is rapidly formed in dense regions, and it is redistributed to diffuse conditions \citep[e.g.][]{Glover_2007b,Micic_2012,Valdivia_2016}. See also Appendix \ref{app:turbulenceH2} for details on the effect of turbulence on H$_2$ abundance.

\subsection{Condition of quasi-steady-state molecular formation}\label{sec:dynamics_static}

\begin{figure*}
	\includegraphics[width=2.0\columnwidth]{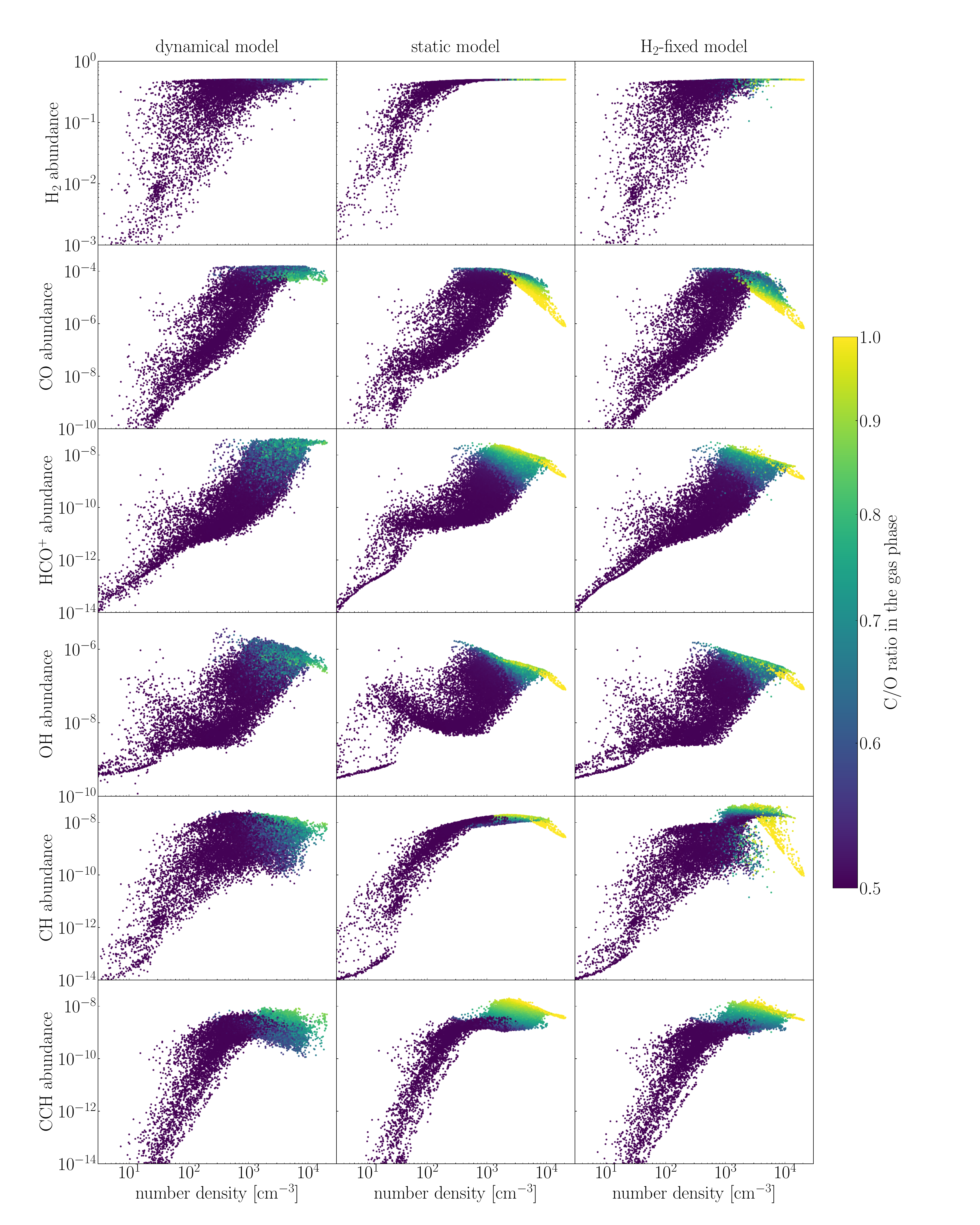}
    \caption{Molecular abundances of H$_2$, CO, HCO$^+$ OH, CH, and CCH relative to hydrogen nuclei in the dynamical model (left column), static model (central column), and H$_2$-fixed model (right column). The horizontal axis shows the gas number density at 5\,Myr. The color scale shows the elemental C/O ratio in the gas phase of each tracer particle.}
    \label{fig:model_summary}
\end{figure*}

\begin{figure*}
	\includegraphics[width=2.0\columnwidth]{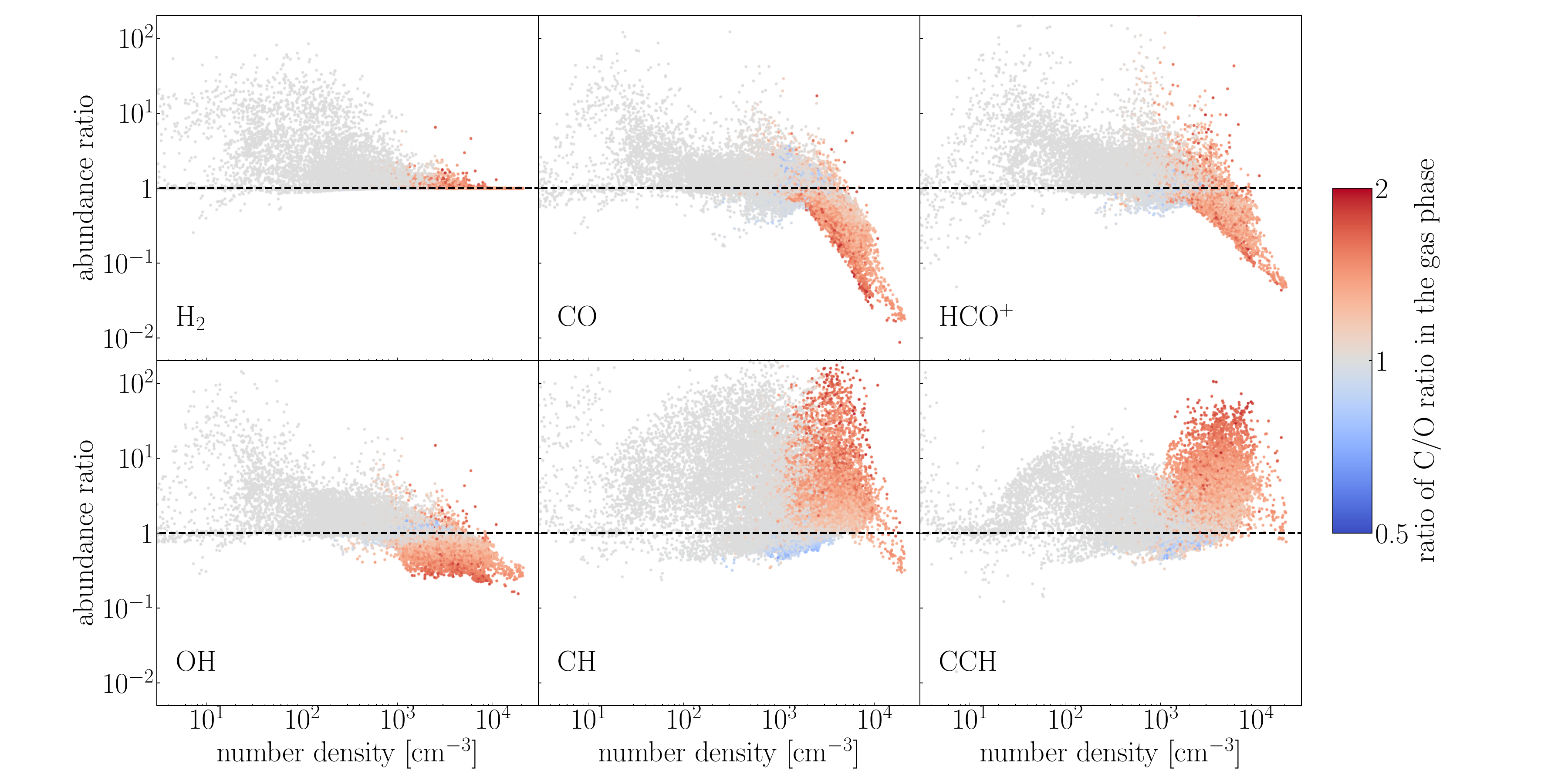}
    \caption{Molecular abundances of static model with respect to those of dynamical model. The color scale shows the gas-phase C/O ratio of the static model divided by that of the dynamical model.}
    \label{fig:dvsp_summary}
\end{figure*}

\begin{figure*}
	\includegraphics[width=2.0\columnwidth]{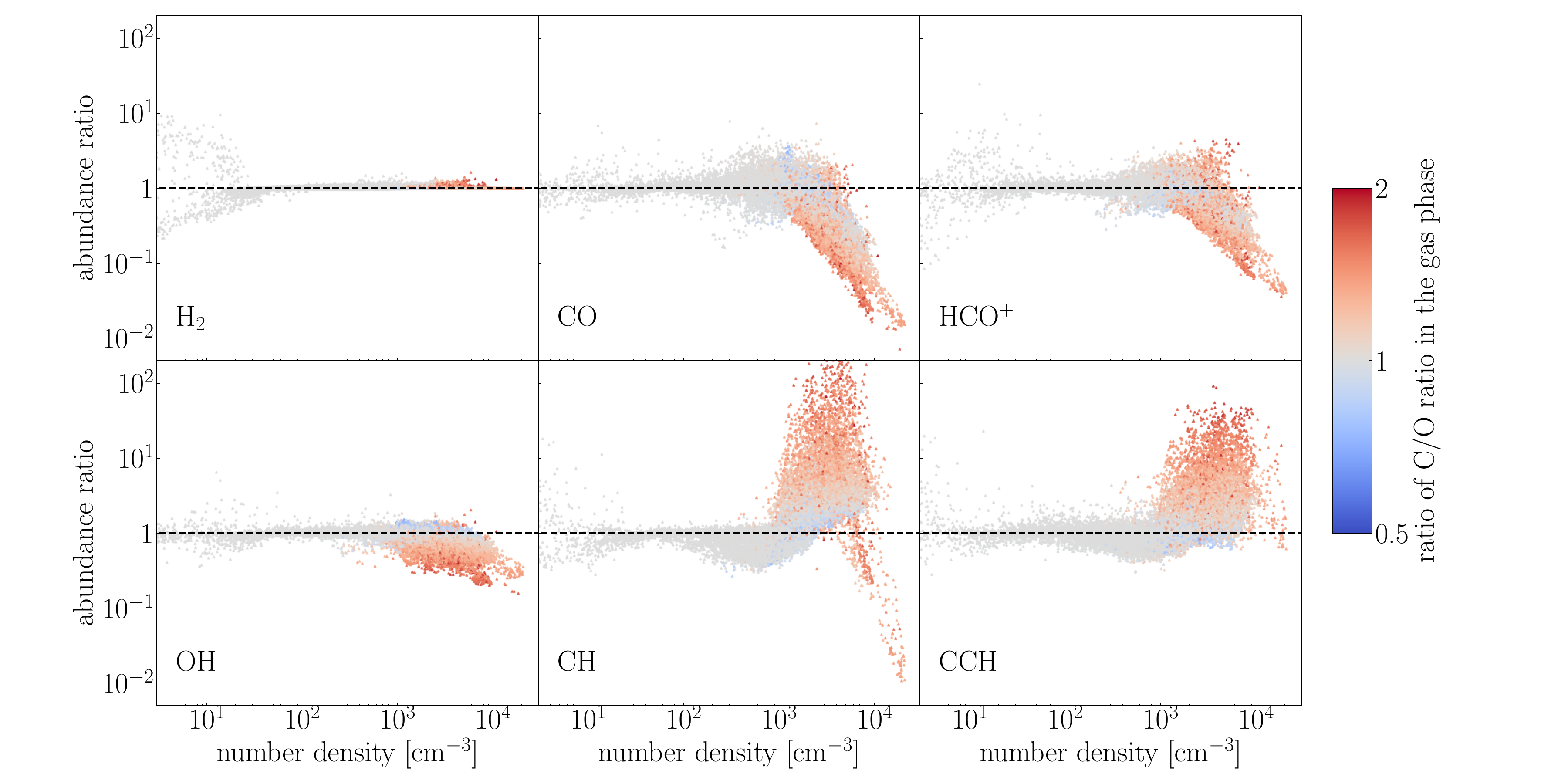}
    \caption{Similar to Figure \ref{fig:dvsp_summary}, but for the H$_2$-fixed model.}
    \label{fig:dvss_summary}
\end{figure*}

\begin{figure*}
	\includegraphics[width=2.0\columnwidth]{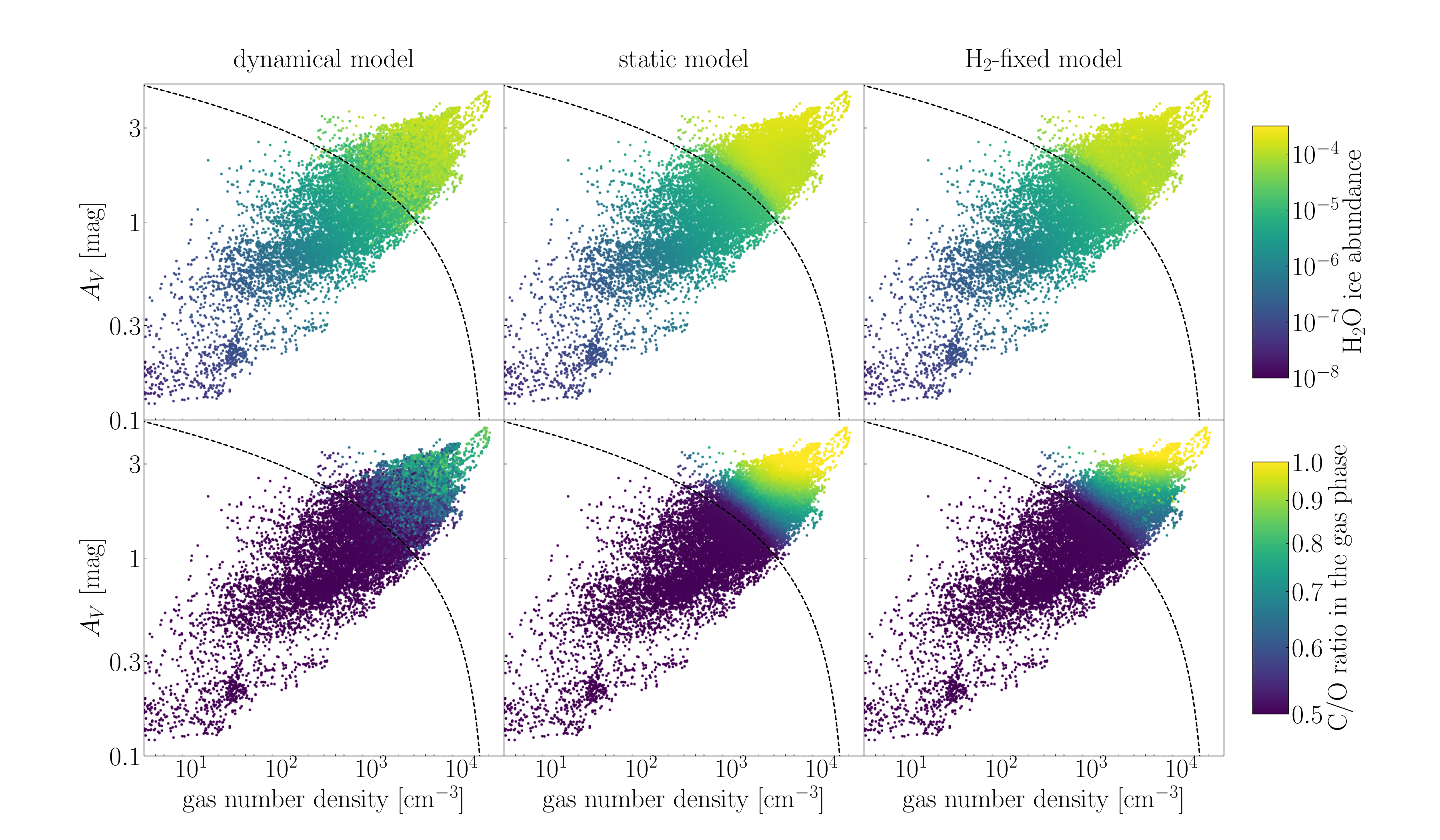}
    \caption{The upper panels show the gas number density, visual extinction, and the H$_2$O ice abundance of each tracer particle at 5\,Myr in the dynamical model (left panel), the static model (center panel), and the H$_2$-fixed model (right panel). The black dashed line shows the threshold visual extinction for H$_2$O ice formation. The lower panels are similar to the upper ones, but for the gas-phase C/O ratio.}
    \label{fig:H2O_COR}
\end{figure*}

Figure \ref{fig:model_summary} shows the gas number density and molecular abundances of dynamical, static, and H$_2$-fixed models. We chose H$_2$ and CO as major molecules, OH as a representative oxygen-bearing molecule, CH and CCH as typical carbon-chain molecules, and HCO$^+$ as a molecule that is closely related to CO. The color scale shows the elemental C/O ratio in the gas phase. To compare the results more quantitatively, we also calculate the ratio of the molecular abundances of the static and dynamical models (Figure \ref{fig:dvsp_summary}). The color scale shows the ratio of the elemental C/O ratio of the two models. Figure \ref{fig:dvss_summary} is similar to Figure \ref{fig:dvsp_summary}, but for the H$_2$-fixed and dynamical models. 

\subsubsection{H$_2$}
Let us first focus on H$_2$, i.e. the top three panels of Figure \ref{fig:model_summary}. The physical conditions and the initial abundances of H$_2$ in the static model are set to be the same as those of the dynamical model at the final timestep. Since H$_2$ formation continues over 5 Myrs, H$_2$ abundances in the static model are generally higher than those of the dynamical model. At $n_\mathrm{H}<10^3\,\mathrm{cm^{-3}}$, H$_2$ abundances of the static model are several to tens of times higher than those of the dynamical model (Figure \ref{fig:dvsp_summary}). At $n_\mathrm{H}>10^3\,\mathrm{cm^{-3}}$, the H$_2$ abundance ratio of almost all particles is close to unity due to the short H$_2$ formation timescale in dense conditions. In the H$_2$-fixed model, on the other hand, the H$_2$ abundance is similar to that of the dynamical model, because the H$_2$ formation rate and the destruction rate are set to be almost equal in the H$_2$-fixed model (Section \ref{sec:ChemModel}). The H$_2$ abundance ratio between the H$_2$-fixed model and dynamical model is almost unity at $n_\mathrm{H}>10\,\mathrm{cm^{-3}}$ (Figure \ref{fig:dvss_summary}). The H$_2$ abundance ratio of tracer particles with $n_\mathrm{H}<10\,\mathrm{cm^{-3}}$ deviates from unity, because some H$_2$ formation routes other than grain surface reactions, such as H$^-$ + H $\rightarrow$ H$_2$ + e$^-$, become dominant. Most of these tracer particles have lower H$_2$ abundance than $10^{-3}$ and hydrogen is almost completely atomic. Since abundances of other molecules should be very low in these particles, we will neglect such particles in this discussion.

\subsubsection{low-density region: $n_{\rm H}\le 10^3$ cm$^{-3}$}
The trends for other molecules are different for diffuse and dense conditions. Firstly, we will focus on the case with gas densities below $10^3\,\mathrm{cm^{-3}}$. In Figure \ref{fig:model_summary}-\ref{fig:dvss_summary}, we can see that for other molecules than H$_2$, the abundances in the H$_2$-fixed model agree well with the dynamical model, compared to those in the static model. The abundance ratios of the H$_2$-fixed and dynamical models for CO, OH, and CH are close to unity (Figure \ref{fig:dvss_summary}). A similar trend is found in HCO$^+$ (upper right panels of Figure \ref{fig:dvss_summary}), which is chemically related to OH and CO, and in CCH (bottom right panels of Figure \ref{fig:dvss_summary}), which is mainly formed via CH (see also Section \ref{sec:each_molecule}). Similar trends are found in other simple molecules not shown in the figures, such as cyanides (CN, HCN).

This result suggests one important property of chemical evolution in diffuse molecular clouds; the molecular abundances of the dynamical model are determined not only by the temporal physical conditions but also by the temporal abundance of H$_2$. The physical conditions of the static model are the same as those of the final timestep of the dynamical models and are kept constant for 5\,Myr. As we mentioned in Section \ref{sec:ChemModel}, the chemistry should reach a steady-state in a shorter timescale than the dynamical timescale when the visual extinction is low enough that photodissociation is effective. Therefore, if the molecular abundances of the dynamical model were determined solely by the steady-state chemistry at fixed physical conditions, the molecular abundances of the dynamical model and the static model should match, but they do not (Figure \ref{fig:dvsp_summary}). On the other hand, the H$_2$-fixed model, whose H$_2$ abundance is consistent with the dynamical model, shows reasonable agreement with the dynamical model. The only difference between the static and H$_2$-fixed models is the H$_2$ abundance. Hence, the molecular abundances of the dynamical model reflect both their temporal physical conditions and H$_2$ abundance. Since H$_2$ is necessary to initiate the formation of the other molecules listed in Figure \ref{fig:dvsp_summary} and \ref{fig:dvss_summary}, it is natural that the H$_2$-fixed model can reproduce the molecular abundances of the dynamical model better than the static model. 
In the static model, the abundances of H$_2$ and other molecules tend to be higher than in the dynamical model.

This finding also suggests that the abundances of simple oxygen-bearing and hydrocarbon molecules in diffuse regions can be well reproduced by steady-state chemistry, given the local physical conditions and H$_2$ abundance. Therefore, molecular abundances in the early stages of molecular cloud evolution can be estimated using hydrodynamical simulations combined with steady-state chemistry, provided that the simulations include a reduced chemical network for H$_2$ and CO that can evaluate the abundances of H$_2$ and the main carbon reservoirs (C$^+$, C, and CO).


\subsubsection{high-density region: $n_{\rm H}> 10^3$ cm$^{-3}$}\label{sec:highdense}
Unlike the diffuse conditions we discussed above, both the static and H$_2$-fixed models do not reproduce the molecular abundances of the dynamical model at $n_\mathrm{H}> 10^3\,\mathrm{cm^{-3}}$ (Figure \ref{fig:dvsp_summary} and \ref{fig:dvss_summary}). CH and CCH abundances of the static and H$_2$-fixed models tend to be higher than in the dynamical model at $10^3\,\mathrm{cm^{-3}}\lesssim n_\mathrm{H}\lesssim  10^4\,\mathrm{cm^{-3}}$, whereas the CO, OH, and HCO$^+$ abundances are lower than in the dynamical model at $n_\mathrm{H}\gtrsim 10^3\,\mathrm{cm^{-3}}$.

This trend comes from the freeze-out of gas-phase chemical species, which changes the elemental C/O ratio in the gas phase. Figure \ref{fig:model_summary} shows that the static and H$_2$-fixed models for high-density region ($n_{\rm H}> 10^3$ cm$^{-3}$) have a larger C/O ratio than the dynamical model. Figure \ref{fig:dvss_summary} also shows that the tracer particles for which molecular abundances differ between the dynamical and H$_2$-fixed models also exhibit differences in their C/O ratios. 

As oxygen atoms are converted to H$_2$O ice, oxygen-bearing molecules such as OH start to decrease in the gas phase (Figure \ref{fig:model_summary}). At the same time, the C/O ratio in the gas phase increases and CO formation slows down. Hydrocarbons and carbon chains are formed from the excess carbon. In addition, CO adsorbs onto the grain surface and forms CO and CO$_2$ ices, which results in the decline of HCO$^+$ abundance. In the case of the static and H$_2$-fixed models, the particles in dense regions continuously experience high-density conditions. It results in efficient freeze-out of oxygen atoms and CO and a significant increase in the C/O ratio in the gas phase compared to the dynamical model (Figure \ref{fig:dvsp_summary} and \ref{fig:dvss_summary}). Note that CH of the H$_2$-fixed models is depleted at $n_\mathrm{H} \sim 10^{4}\,\mathrm{cm^{-3}}$, because the abundance of H atoms, which is the main reactant destroying CH (see also Section \ref{sec:CH_CCH}), is enhanced as a result of the modified desorption products in the model (see Section \ref{sec:ChemModel}).

The critical condition for the increase in the C/O ratio and for the difference between the dynamical and the static model (or the H$_2$-fixed model) corresponds to the threshold visual extinction for H$_2$O ice formation, $A_V^{\mathrm{th}}$, proposed by \citet{Hollenbach_2009} \citep[see also][]{Tielens_2005}. H$_2$O ice formation is possible when the adsorption of oxygen atoms proceeds faster than the photodesorption of H$_2$O ice. $A_{V}^{\mathrm{th}}$ is given as follows: 
\begin{equation}
    A_{V}^{\mathrm{th}}=0.56\ln\,\left[\frac{Y_\mathrm{H_2O}G_0F_0}{c_\mathrm{s}A_\mathrm{O}n_\mathrm{H}}\right],
\end{equation}
where $F_0=10^8\,\mathrm{photon\,cm^{-2}\,s^{-1}}$ is the flux of the FUV photons with $G_0=1$, $c_\mathrm{s}$ is the sound speed of an oxygen atom, $A_\mathrm{O}=3.2\times 10^{-4}$ is the elemental abundance of oxygen. $Y_\mathrm{H_2O}=1\times 10^{-3}$ is the photodesorption yield of H$_2$O from H$_2$O ice (see Section \ref{sec:ChemSetup}). 

Figure \ref{fig:H2O_COR} shows the H$_2$O ice abundance and the C/O ratio in the gas phase on the $n_\mathrm{H}$-$A_V$ plane. The dashed line shows the threshold visual extinction, $A_{V}^{\mathrm{th}}$, above which H$_2$O ice is well formed. Especially, the H$_2$O ice abundance in the static and H$_2$-fixed models sharply changes at the threshold visual extinction. In the dynamical model, the contrast is less clear due to the temporal variation of the gas density and visual extinction in the dynamical model. The gas-phase C/O ratio also increases where the H$_2$O ice formation proceeds (the bottom panels of Figure \ref{fig:H2O_COR}). The typical gas number density at which the C/O ratio starts to increase is about $10^3\,\mathrm{cm^{-3}}$. This value is consistent with the gas density above which the molecular abundances in the dynamical and H$_2$-fixed models deviate (Figure \ref{fig:dvss_summary}).

Our previous work \citep{Komichi_2024} studied the chemical evolution during molecular cloud formation by a one-dimensional steady-state shock model. They show similar results; the H$_2$O ice formation affects the abundances of C-bearing or O-bearing molecules. It is noteworthy that the trend of the fiducial model in \citet{Komichi_2024} is close to that of the static and H$_2$-fixed models in this work, not the dynamical model. Since the visual extinction of the steady-state shock model gradually increases, the fluid element experiences dense conditions for a timescale much longer than the freeze-out timescale, resulting in a decrease of the CO abundance by about one order of magnitude. In contrast, the dynamical model in the present work does not show such strong CO depletion, because tracer particles do not necessarily remain in dense regions. \citet{Panessa_2023} and \citet{Priestley_2023} also investigated the chemical evolution of molecular clouds using tracer particles, and found that CO depletion does not proceed significantly at densities of $n_\mathrm{H}\sim 10^4\,\mathrm{cm^{-3}}$.

\section{Discussion}\label{sec:discussion}

\subsection{CO abundance and effect of CO self-shielding factor}\label{sec:CO}

\begin{figure}
	\includegraphics[width=\columnwidth]{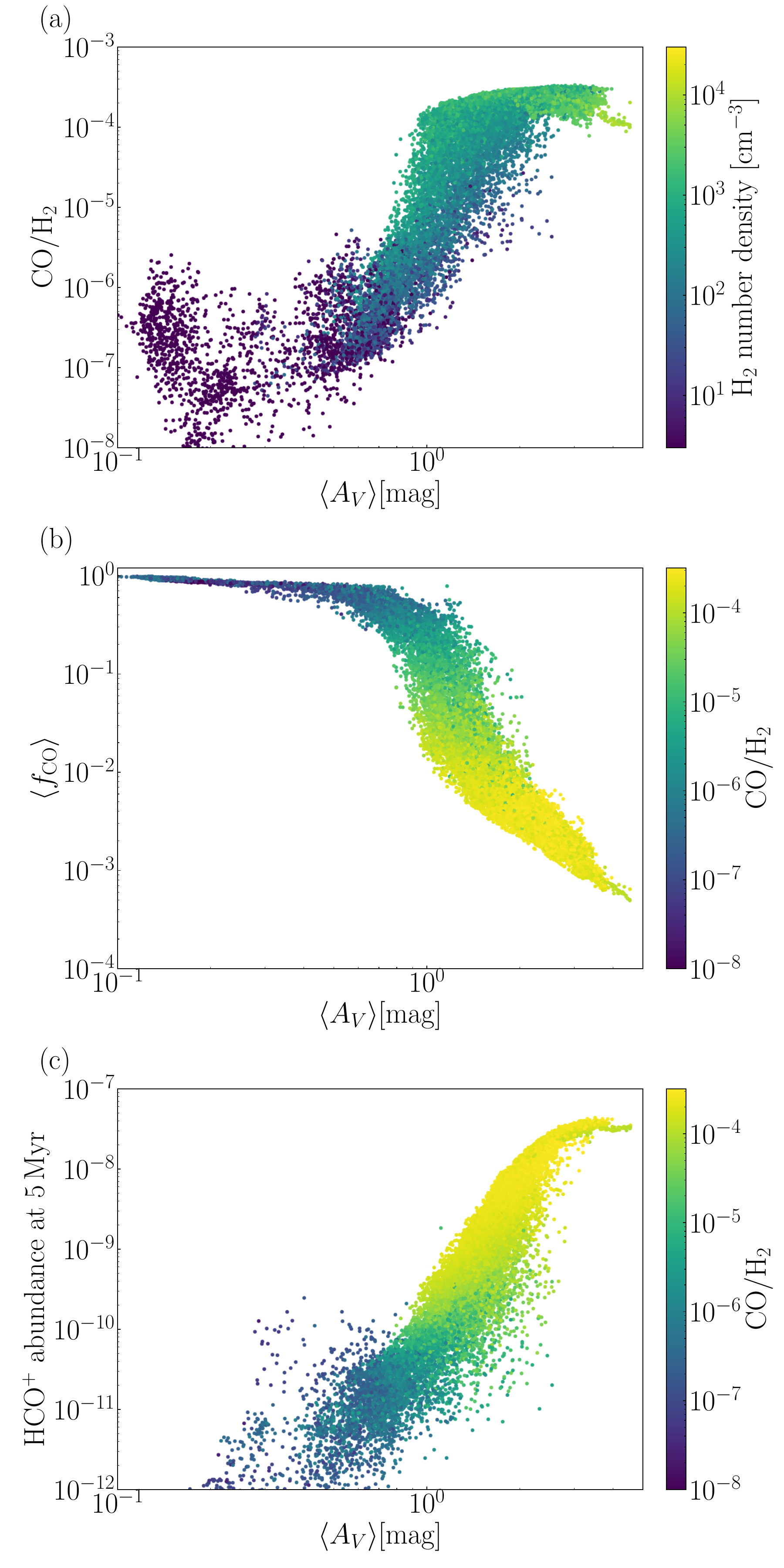}
    \caption{(a) The effective visual extinction and the CO abundance relative to H$_2$ abundance. The color scale shows the H$_2$ number density. (b) The effective visual extinction and the effective self and mutual shielding factor of each tracer particle. The color scale shows the CO abundance relative to H$_2$ abundance. (c) The effective visual extinction and HCO$^+$ abundance relative to hydrogen nuclei of each tracer particle. The color scale is the same as that of panel (c).}
    \label{fig:CO_HCOp}
\end{figure}

First, we examine the abundance of CO relative to H$_2$. Figure \ref{fig:CO_HCOp} (a) shows the effective visual extinction and CO abundance. The color scale indicates the H$_2$ number density. The CO abundance approximately saturates at $\langle A_V\rangle\gtrsim 3\,\mathrm{mag}$, while the CO abundance exhibits a scatter of more than one order of magnitude at $\langle A_V\rangle\lesssim 3\,\mathrm{mag}$. We can also see that the CO abundance varies significantly among the tracer particles with a similar H$_2$ number density. These indicate that neither gas density nor visual extinction determines the local CO abundance, as also suggested by \citet{Glover_2010}. As discussed below, this behavior can be attributed to the self and mutual-shielding effects of CO, which depend on the spatial distribution of CO number density.

Next, we will focus on the relationship between the CO self-shielding effect and CO abundance. Some previous works conclude that the dust shielding is essential for CO formation without mentioning the self-shielding effect \citep[e.g.][]{Inoue_2012,Gong_2018,Iwasaki_2019}. Here we verify this point. In diffuse molecular clouds, CO is mainly destroyed by the external radiation field through photodissociation. The photodissociation rate is controlled by the UV absorption by dust, self-shielding, and mutual shielding by H$_2$. If the self and mutual shielding are neglected, the attenuation factor of the photodissociation by dust is given as follows:
\begin{equation}
    \langle \exp[-\beta(\mathrm{CO})A_V]\rangle = \frac{1}{6}\sum_{\mathrm{i}=\pm x,\pm y,\pm z} \exp(-\beta(\mathrm{CO})A_{V,\mathrm{i}}),
\end{equation}
where $\beta(\mathrm{CO})$ is a coefficient for CO shielding by dust grains. When the self-shielding and the mutual shielding by H$_2$ are included, the attenuation factor is rewritten as below:
\begin{equation}
    \begin{split}
        \langle f_\mathrm{CO}\exp[-\beta(\mathrm{CO})A_V]\rangle = \frac{1}{6}\sum_{\mathrm{i}=\pm x,\pm y,\pm z} & f_\mathrm{CO}(N_\mathrm{CO,i},N_\mathrm{H_2,i})\\
        &\exp(-\beta(\mathrm{CO})A_{V,\mathrm{i}}),
    \end{split}
\end{equation}
where $f_\mathrm{CO}$ is the self and mutual shielding factor of CO, and $N_\mathrm{CO,i}$ is the column density of CO calculated by integrating the CO number density from the i boundary. Then the effective self and mutual shielding factor, $\langle f_\mathrm{CO} \rangle$ is defined as follows:
\begin{equation}
    \langle f_\mathrm{CO} \rangle=\frac{\langle f_\mathrm{CO}\exp[-c(\mathrm{CO})A_V]\rangle}{\langle \exp[-c(\mathrm{CO})A_V]\rangle}.
\end{equation}
If this value is much smaller than unity, it means that the self and mutual shielding dominate the attenuation of CO. Figure \ref{fig:CO_HCOp} (b) shows the visual extinction and $\langle f_\mathrm{CO} \rangle$ of the tracer particles. The color scale shows the CO abundance in the gas phase. While the CO abundance starts to increase around $A_V\simeq 1\,\mathrm{mag}$, where the shielding by dust grains becomes important, the CO abundance saturates when $\langle f_\mathrm{CO} \rangle$ gets lower than $10^{-2}$, i.e., the self and mutual shielding become effective. The CO abundance at $A_V\gtrsim 1\,\mathrm{mag}$ depends on $\langle f_\mathrm{CO} \rangle$ rather than the visual extinction, suggesting that the self and mutual shielding are essential for completing CO formation, although the dust shielding is important for initiating the CO formation. 

\subsection{Major reactions and analytic solutions}\label{sec:each_molecule}

\begin{figure}
	\includegraphics[width=1.0\columnwidth]{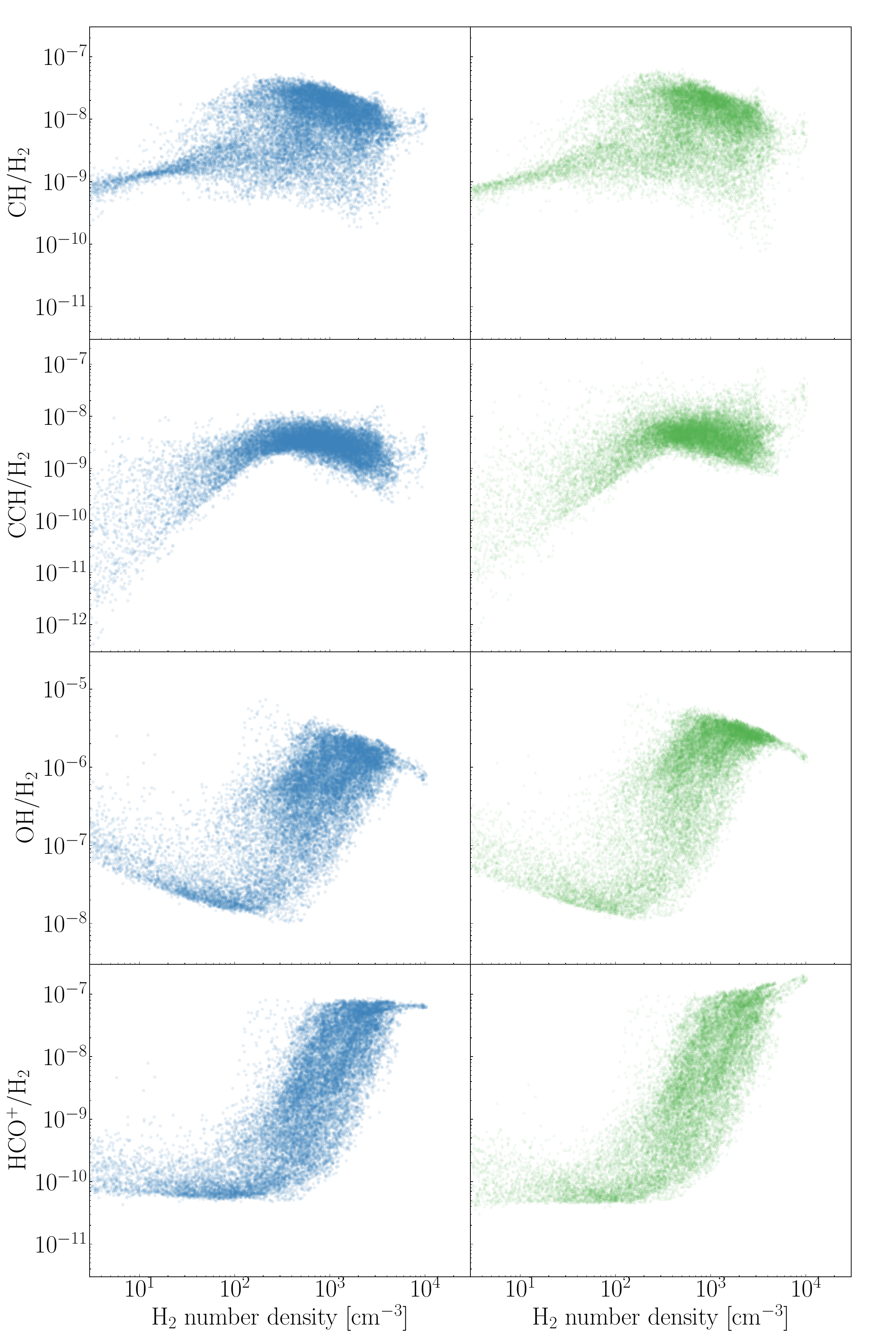}
    \caption{Left column: H$_2$ number density and the molecular abundance relative to that of H$_2$ of the dynamical model. Right column: similar to the left panels, but the molecular abundances are calculated with the analytic solutions we derive in Appendix \ref{app:analyticsol}.}
    \label{fig:CH_CCH_OH_HCOp}
\end{figure}

Here we look into the formation and destruction paths of several molecules relevant to the observation of diffuse molecular clouds. The chemical timescale of radicals such as CH, CCH, and OH is short, especially in diffuse conditions where the photodissociation is effective. As discussed in detail below, their abundances can be estimated by analytic solutions provided the abundances of major species (e.g., H$_2$ and major carbon carriers) are determined by e.g. the simple reaction network implemented in the MHD simulations. We describe the major reactions included in the analytic solutions in Section \ref{sec:CH_CCH} and \ref{sec:OH_HCOp}, and compare the analytic solutions with numerical results in Section \ref{sec:ans_ab}. The derivation of the analytic solutions is described in Appendix \ref{app:analyticsol}.

\subsubsection{CH and CCH} \label{sec:CH_CCH}

The top left panel of Figure \ref{fig:CH_CCH_OH_HCOp} shows the H$_2$ number density and CH abundance at the final time step (i.e. $t=5\,\mathrm{Myr}$) in the dynamical model. CH formation is initiated by the two-body reactions C$^+$ + H$_2$ $\rightarrow$ CH$_2^+$ and C + H$_3^+$ $\rightarrow$ CH$^+$ + H$_2$, followed by reactions with H$_2$ to form CH$_3^+$. CH$_2^+$ and CH$_3^+$ recombine with electrons to form CH or CH$_2$. The latter can also be converted to CH by a reaction with H atom.

CH abundance shows a scatter at $n(\mathrm{H_2)}\gtrsim 10^2\,\mathrm{cm^{-3}}$. There are two main reasons for this. One is the variation in temporal physical conditions, such as gas density and visual extinction. Their variations especially affect the abundances of C$^+$ and C, which are closely related to CH formation. The other is the fraction of atomic hydrogen, which is the major reactant of CH: CH + H $\rightarrow$ C + H$_2$. H atom abundance, as well as H$_2$ abundance, depends on the physical history of each fluid element as discussed in Section \ref{sec:H2}. Therefore, particles with the similar H$_2$ number density at the final time step have different H atom abundance, which results in a variety of CH abundance. Based on these major reactions, the CH abundance is analytically described by equation (\ref{eq:CH}).

Many particles with the final $n(\mathrm{H_2})>10^2\,\mathrm{cm^{-3}}$ have the highest CH/H$_2$ value around $10^{-8}$. These have a low atomic hydrogen number density of around $10\,\mathrm{cm^{-3}}$ at the final time step, which is the lower limit value inferred from equation (\ref{eq:H2analytic}), where the H$_2$ formation and the destruction of H$_2$ by cosmic rays are balanced. This means that if fluid elements are well self-shielded by H$_2$ and experience high-density conditions for a long enough time, the CH abundance relative to H$_2$ is saturated with almost a constant value. 


The second row of Figure \ref{fig:CH_CCH_OH_HCOp} shows the CCH abundance relative to H$_2$. There are mainly three pathways for starting CCH formation. One is the ion-neutral reaction C$^+$ + CH $\rightarrow$ C$_2^+$ + H followed by C$_2^+$ + H$_2$ $\rightarrow$ C$_2$H$^+$ + H and C$_2$H$^+$ + H$_2$ $\rightarrow$ C$_2$H$_2^+$ + H. This is efficient when C$^+$ is the major carbon carrier. In addition, C$^+$ + CH$_4$ $\rightarrow$ C$_2$H$_2^+$ + H$_2$ (or C$_2$H$_3^+$ + H) can be effective if CH$_4$ is produced via non-thermal desorption of icy CH$_4$ on grain surfaces. If atomic carbon is abundant, another two-body reaction can initiate CCH formation: CH$_3^+$ + C $\rightarrow$ C$_2$H$_2^+$ + H. Once C$_2$H$_2^+$ is formed, the electron recombination C$_2$H$_2^+$ + e$^-$ $\rightarrow$ CCH + H can produce CCH. Besides this reaction, some other hydrocarbons which are closely related to C$_2$H$_2^+$ can also help. For instance, C$_2$H$_4^+$ is directly formed by the radiative association C$_2$H$_2^+$ + H$_2$ $\rightarrow$  C$_2$H$_4^+$. This is destroyed by atomic hydrogen, which results in C$_2$H$_3^+$ (C$_2$H$_4^+$ + H $\rightarrow$ C$_2$H$_3^+$ + H$_2$). Similar to C$_2$H$_2^+$, the recombination of C$_2$H$_3^+$ with an electron produces CCH. In addition, the electron recombination reactions of C$_2$H$_3^+$ and C$_2$H$_4^+$ also form C$_2$H$_2$, and its destruction by photodissociation can produce CCH. 

CH$_4$ formation on grain surfaces is still under discussion. The infrared observations have shown that CH$_4$ ice is not abundant \citep{Oberg_2008, McClure_2023}, whereas the astrochemical models tend to overproduce CH$_4$ ice \citep[e.g.][]{Garrod_2006}. Recently, \citet{Tsuge_2024} revealed that CH$_4$ formation on H$_2$O ice is inefficient. However, in the cloud formation condition, other ices like NH$_3$ are also formed, and CH$_4$ formation efficiency on such ices is still not well known. Recently, \citet{Molpeceres_2024} found that adsorbed carbon atoms on NH$_3$ ice can be converted to some molecules like CH$_3$NH$_2$, HCN, and HNC, suggesting that CH$_4$ ice formation is inefficient on NH$_3$ ice. We emphasize that the understanding of CH$_4$ ice formation and its non-thermal desorption on various surface conditions is required.

\subsubsection{OH and HCO$^+$} \label{sec:OH_HCOp}

The major formation route of OH and HCO$^+$ is similar to that presented in \citet{Komichi_2024} (see Panel (a) of Figure 5 in \citet{Komichi_2024}). Once H$_2$ and H$_3^+$ are formed, OH formation starts from H$_3^+$ + O $\rightarrow$ OH$^+$ + H$_2$. Through a series of reaction with H$_2$, OH$^+$ is converted to H$_3$O$^+$, which recombines with an electron to form OH. In addition, the photodesorption of H$_2$O from the grain surfaces also helps the OH formation. The destruction route of OH depends on the physical conditions. In diffuse conditions, the photodissociation and the two-body reactions with ions such as H$^+$ and C$^+$ are the main routes. In dense conditions where the abundances of such ions decline, the two-body reactions with neutral species like O and S become effective. 

In diffuse conditions where C$^+$ is the major carbon carrier, HCO$^+$ formation is initiated by C$^+$ + OH $\rightarrow$ CO$^+$ + H. Then CO$^+$ reacts with H$_2$ as CO$^+$ + H$_2$ $\rightarrow$ HCO$^+$ + H. In addition, similar to OH, the grain surface reactions also help through the desorption of H$_2$O; H$_2$O in the gas phase also reacts with C$^+$ and forms HCO$^+$ by C$^+$ + H$_2$O $\rightarrow$ HCO$^+$ + H. On the other hand, once CO becomes abundant, the two-body reaction H$_3^+$ + CO $\rightarrow$ HCO$^+$ + H$_2$ dominates the HCO$^+$ formation. This trend can be found in Figure \ref{fig:CO_HCOp} (c); most particles with the HCO$^+$ abundance higher than 10$^{-10}$ show abundant CO.


\subsubsection{Analytic solution of molecular abundances}\label{sec:ans_ab}

The right column of Figure \ref{fig:CH_CCH_OH_HCOp} shows the analytic solutions of the molecular abundances assuming steady-state conditions. We construct the analytic solutions based on the major reactions that are important in diffuse or dense conditions as we discussed above. The input parameters are physical quantities ($n_\mathrm{H}, T_\mathrm{gas},$ and $A_V$) and the abundances of major species (H, H$_2$, H$^+$, H$_3^+$, C$^+$, C, CO, O, and e$^-$), which are also included in the chemical network used in our MHD simulations (see Section \ref{sec:MHDequ}). 
The number densities of CH, CCH, OH, and HCO$^+$ are expressed as follows:
\begin{equation}\label{eq:nCH}
    \begin{split}
        &n(\mathrm{CH})\\
        =&\frac{\left(f_\mathrm{CH_3^++e^-}^\mathrm{CH}+f_\mathrm{CH_3^++e^-}^\mathrm{CH_2}\right)k_\mathrm{CH_2^++H_2}n(\mathrm{H_2})+f_\mathrm{CH_2^++e^-}^\mathrm{CH}k_\mathrm{CH_2^++e^-}n(\mathrm{e^-})}{k_\mathrm{CH_2^++H_2}n(\mathrm{H_2})+k_\mathrm{CH_2^++e^-}n(\mathrm{e^-})}\\
        &\times\frac{k_\mathrm{C^++H_2}n(\mathrm{C^+})n(\mathrm{H_2})+k_\mathrm{C+H_3^+}n(\mathrm{C})n(\mathrm{H_3^+})}{k_\mathrm{CH+H}n(\mathrm{H})+k_\mathrm{CH,pd}},
    \end{split}
\end{equation}
\begin{equation}\label{eq:nCCH}
    n(\mathrm{CCH})=\frac{F(\mathrm{CCH})}{D(\mathrm{CCH})},
\end{equation}
where
\begin{equation}\label{eq:F_CCH}
    \begin{split}
        &F(\mathrm{CCH})\\&=f_\mathrm{C_2H_2^++e^-}^\mathrm{CCH}k_\mathrm{C_2H_2^++e^-}n(\mathrm{C_2H_2^+})n(\mathrm{e^-})\\&+f_\mathrm{C_2H_3^++e^-}^\mathrm{CCH}k_\mathrm{C_2H_3^++e^-}n(\mathrm{C_2H_3^+})n(\mathrm{e^-})+k_\mathrm{C_2H_2,pd}n(\mathrm{C_2H_2}),
    \end{split}
\end{equation}
and,
\begin{equation}\label{eq:D_CCH}
    D(\mathrm{CCH})=k_\mathrm{CCH+C^+}n(\mathrm{C^+})+k_\mathrm{CCH+O}n(\mathrm{O})+k_\mathrm{CCH,pd},
\end{equation}
\begin{equation}\label{eq:nOH}
    n(\mathrm{OH})=\frac{K_1K_2+K_3}{K_4-K_1K_5},
\end{equation}
where
\begin{equation}\label{eq:OHcomp}
    \begin{split}
        &K_1=\frac{f_\mathrm{H_3O^++e^-}^\mathrm{OH}k_\mathrm{H_2O^++H_2}n(\mathrm{H_2})}{k_\mathrm{H_2O^++H_2}n(\mathrm{H_2})+k_\mathrm{H_2O^++e^-}n(\mathrm{e^-})},\\
        &K_2=k_\mathrm{O^++H_2}n(\mathrm{O^+})n(\mathrm{H_2})+k_\mathrm{H_3^++O}n(\mathrm{H_3^+})n(\mathrm{O}),\\
        &K_3=p_\mathrm{O}k_\mathrm{O, dust}n(\mathrm{O}),\\
        &K_4=k_\mathrm{C^++OH}n(\mathrm{C^+})+k_\mathrm{H^++OH}n(\mathrm{H^+})+k_\mathrm{O+OH}n(\mathrm{O})\\&+k_\mathrm{S+OH}n(\mathrm{S})+k_\mathrm{H_3^++OH}n(\mathrm{H_3^+})+k_\mathrm{OH,pd},\\
        &\mathrm{and,}\\
        &K_5=k_\mathrm{H_3^++OH}n(\mathrm{H_3^+}),
    \end{split}
\end{equation}
and, 
\begin{equation}\label{eq:nHCOp}
    n(\mathrm{HCO^+})=\frac{F(\mathrm{HCO^+})}{D(\mathrm{HCO^+})},
\end{equation}
where
\begin{equation}\label{eq:F_HCOp}
    \begin{split}
        &F(\mathrm{HCO^+})\\&=f_\mathrm{CO^++H_2}^\mathrm{HCO^+}k_\mathrm{CO^++H_2}n(\mathrm{CO^+})n(\mathrm{H_2})+k_\mathrm{H_3^++CO}n(\mathrm{H_3^+})n(\mathrm{CO})\\&+f_\mathrm{H_3^++CO}^\mathrm{HCO^+}k_\mathrm{C^++H_2O}n(\mathrm{C^+})n(\mathrm{H_2O}),
    \end{split}
\end{equation}
and,
\begin{equation}\label{eq:D_HCOp}
    D(\mathrm{HCO^+})=k_\mathrm{HCO^++e^-}n(\mathrm{HCO^+})n(\mathrm{e^-}).
\end{equation}
The definition of the symbols and the derivation of equation (\ref{eq:nCH})-(\ref{eq:D_HCOp}) and molecular abundances (other than input parameters) in the equations are described in Appendix \ref{app:analyticsol}.
Here we calculated the molecular abundances using the physical quantities of tracer particles and the abundances of major species from the dynamical model. The calculated molecular abundances with the analytic solutions show reasonable agreement with the numerical results, both in diffuse and dense conditions. In dense regions, molecular abundances can be reproduced by our analytic solutions if we know the abundances of major species determined by grain surface chemistry (e.g., H$_2$O ice formation).

In Section \ref{sec:highdense}, we discussed the temporal variation of the gas-phase C/O ratio due to water ice formation. Recalling this point, the agreement of the abundances in the dynamical model and analytical solutions has two implications. First, molecular abundances of minor species evolve in a non-equilibrium way through the time variation of the abundances of major molecules and atoms in the gas phase. Second, once the abundances of major chemical species are given, the abundances of radicals are mostly determined with steady-state chemistry in the cloud formation phase. This behavior is expected, given the short chemical timescale of radicals (e.g., CH, CCH, and OH), especially in diffuse conditions ($n_\mathrm{H}<10^3\,\mathrm{cm^{-3}}$), where the visual extinction is still below $1-3\,\mathrm{mag}$ and photodissociation is effective (see Figure \ref{fig:phys_tracer}, \ref{fig:dvsp_summary}, or \ref{fig:dvss_summary}). We also confirm that the molecular abundances of the static and H$_2$-fixed models can be reproduced with the analytic solutions unless formation routes that are not considered in the analysis (e.g., grain surface chemistry) become dominant (see Appendix \ref{app:analyticsol}).

\subsection{Comparison with observations of diffuse molecular clouds}\label{sec:vs_obs}

\begin{figure}
	\includegraphics[width=1.0\columnwidth]{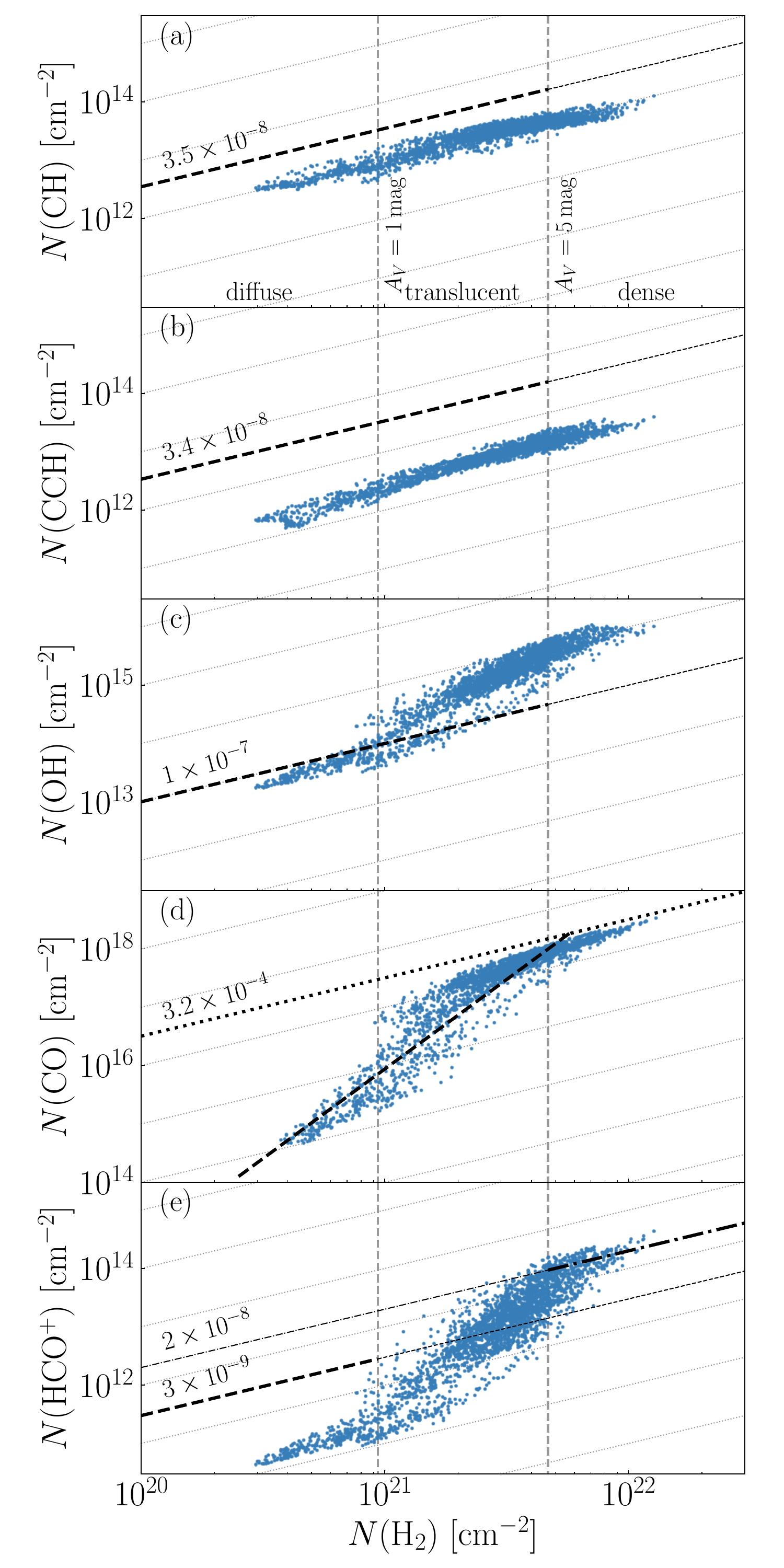}
    \caption{The blue dots show the calculated column densities of H$_2$, CH, CCH, OH, CO, and HCO$^+$ of each tracer particle. The vertical gray dashed lines show the H$_2$ column densities corresponding to $A_V=1\,\mathrm{mag}$ and $5\,\mathrm{mag}$. The black dashed lines show the typical abundance of CH \citep[$3.5\times 10^{-8}$;][]{Sheffer_2008}, CCH \citep[$3.4\times 10^{-8}$;][]{Thiel_2019}, OH \citep[$1\times 10^{-7}$;][]{Lucas_1996,Jacob_2022}, CO \citep[see the main text;][]{Sheffer_2008}, and HCO$^+$ in diffuse clouds\citep[$3\times 10^{-9}$;][]{Thiel_2019}. The black dotted line shows the upper limit of the CO abundance relative to H$_2$, $3.2\times 10^{-4}$. The black dot-dashed line shows the typical HCO$^+$ abundance in dense clouds\citep[$2\times 10^{-8}$;][]{Fuente_2019,Thiel_2019}. The black dashed lines are highlighted in diffuse and translucent regions for CH, CCH, and OH, as the observational value is not necessarily valid in dense clouds. For HCO$^+$, the black dashed and dot-dashed lines are highlighted in diffuse and dense regions, respectively, following \citet{Thiel_2019}. The gray dotted lines show the constant ratio of H$_2$ and molecular column densities.}
    \label{fig:remap_mol}
\end{figure}

In this section, we compare our results with observational studies of molecular species in diffuse and translucent clouds, which indicate that the column densities of certain molecules exhibit a linear correlation with that of H$_2$. Based on this relationship, such molecules are often considered as potential tracers of H$_2$ (i.e., molecular gas).
We investigate if and why such linearity appears in our model. While we solve the large chemical network for tracer particles, we can derive molecular column densities using remapping technique. The details of the adopted algorithm are described in Appendix \ref{app:remapping}. As we impose the periodic conditions for $y$ and $z$ boundaries in our MHD simulations, we only focus on the column density along the $x$-axis. When observing the clouds or the compression layers at an oblique angle to the $x$-axis or observing multiple clouds, the column density is simply multiplied by a constant, which does not affect the discussion below.


Blue dots in Figure \ref{fig:remap_mol} show the calculated column densities of H$_2$ and the molecules discussed in Section \ref{sec:CO} and \ref{sec:each_molecule}. The gray dotted lines show the constant ratio of H$_2$ and molecular column densities. 
Panel (a) shows the column density of CH, which shows a linear correlation with H$_2$ column density. The typical column density ratio in our model is $\sim 10^{-8}$, which is consistent with the highest CH abundance at $n(\mathrm{H_2})\sim 10^{2-3}\,\mathrm{cm^{-3}}$ (Figure \ref{fig:CH_CCH_OH_HCOp}). Although there is a scatter of one order of magnitude in CH abundance at those densities, the CH column density shows a scatter of only a factor of a few. As explained in Section \ref{sec:CH_CCH}, CH becomes abundant as H$_2$ fraction increases because CH is destroyed by atomic hydrogen. The tracer particles with $n(\mathrm{H_2})\gtrsim 10^2\,\mathrm{cm^{-3}}$ tend to have a high H$_2$ fraction, and thus a high CH abundance. These regions with high H$_2$ and CH abundances contribute most to their column densities.

The dashed line shows the typical value of the CH column density derived from the absorption line observations in optical and ultraviolet observations \citep{Sheffer_2008}. Our results show reasonable agreement with the observations with CH abundance in terms of linearity, with the CH abundance underestimated only by a factor of three. This suggests that the H$_2$ abundances of interstellar clouds observed by CH absorptions are close to being saturated. As \citet{Bottinelli_2014} points out, the CH abundance sensitively depends on the reaction rate coefficient of CH + H $\rightarrow$ C + H$_2$, which is the major destruction path for CH. In the present work, we adopted the rate coefficient from \citet{vanHarrevelt_2002}, which is recommended by KIDA database\footnote{\url{https://kida.astrochem-tools.org/}} \citep{Wakelam_2012}. If the rate coefficient were three times smaller than the adopted value at $T_\mathrm{gas}\sim 10\,\mathrm{K}$, the calculated CH abundance would agree with the observations. 

Panel (b) shows the CCH column density. Similar to CH, CCH also shows linear correlations with H$_2$ in our model. This result is natural because CCH is mainly form from CH. Therefore, CCH can also be useful for tracing diffuse molecular clouds. Our model underestimates the typical column density ratio $N(\mathrm{CCH})/N(\mathrm{H_2})$ derived from observations of diffuse and translucent clouds \citep{Thiel_2019}. Since the column density ratio of CH/H$_2$ in our model is lower than the observed value by a factor of three, there may be another systematic uncertainty of a factor of three in the rate coefficients along the chemical pathway from CH to CCH, assuming that other sources of errors (e.g., difference in the physical conditions) are negligible. 

Panel (c) shows the column densities of H$_2$ and OH. The dashed line shows the typical column density ratio $N(\mathrm{OH})/N(\mathrm{H_2})$ derived from the observations \citep[e.g.][]{Lucas_1996,Jacob_2022}. The calculated column densities agree with the observational value, especially around $N(\mathrm{H_2})\sim 10^{21}\,\mathrm{cm^{-2}}$. In such diffuse regions, OH is mainly destroyed by photodissociation and by the two-body reaction with C$^+$. As the visual extinction and the gas density increase, photodissociation becomes inefficient and the C$^+$ abundance decreases. Instead, the reactions with neutral atoms such as O and S become the dominant destruction pathways of OH. In parallel, as H$_2$ formation proceeds, H$_3^+$ becomes more abundant, which further contributes to OH formation. As a result, the calculated OH column density shows a slight increase, which is consistent with the observations by \citet{Jacob_2022}; the column density ratio of OH relative to H$_2$ increases by a factor of several to ten as $N(\mathrm{H_2})$ increases from $10^{20}\,\mathrm{cm^{-2}}$ to $10^{22}\,\mathrm{cm^{-2}}$ \citep[see Figure 11 of][]{Jacob_2022}.

Panel (d) shows the CO column density. The calculated CO column density shows a non-linear dependence on H$_2$ column density, which is consistent with the rapid rise of CO/H$_2$ abundance ratio as $A_V$ and gas density increase (Figure \ref{fig:CO_HCOp} a). Our result is consistent with the empirical relation derived from the observations of diffuse and dark clouds above $N(\mathrm{H_2})=20.4\mathrm{cm^{-2}}$: $N(\mathrm{CO})=10^{14.1}\,\mathrm{cm^{-2}}(N(\mathrm{H_2})/10^{20.4}\,\mathrm{cm^{-2}})^{3.07}$ \citep{Sheffer_2008}, which is indicated with the dashed line. As discussed in \citet{Sheffer_2008}, the plane-parallel static model of \citet{vanDishoeck_1988} also reproduces the observed relationship between H$_2$ and CO column densities. Our results demonstrate that this relationship can be reproduced in the time-dependent, three-dimensional simulations, probably because it is mainly determined by the self-shielding effects.

Panel (e) shows that HCO$^+$ column density does not linearly scale with that of H$_2$, and it has a scatter of more than one order of magnitude at a given H$_2$ column density. This is because HCO$^+$ formation path is different in diffuse and dense conditions, reflecting the CO abundance (see Section \ref{sec:OH_HCOp}). The dashed line shows the typical abundance of HCO$^+$ ($3\times 10^{-9}$) derived from diffuse cloud observations, while the dashdot line shows that of the dark cloud TMC1 \citep{Fuente_2019}. \citet{Thiel_2019} point out that the HCO$^+$ abundance in translucent clouds should be between the above two values \citep[see also][]{Kim_2023}. Our results reproduce the HCO$^+$ abundance of the dark cloud at $N(\mathrm{H_2})\gtrsim 5\times 10^{21}\,\mathrm{cm^{-2}}$. At $N(\mathrm{H_2})\lesssim 1\times 10^{21}\,\mathrm{cm^{-2}}$, HCO$^+$ column density in our model is lower than the observation.

There are some possibilities to explain the observational value of HCO$^+$ at low $N(\mathrm{H_2})$. Firstly, each velocity component detected by the observations of molecular absorption lines consists of a mixture of CO-rich and CO-poor clouds. In other words, observed clouds have clumpy CO-rich gas surrounded by CO-poor diffuse gas. Since the HCO$^+$ abundance non-linearly increases with gas density, the column density can be enhanced due to locally HCO$^+$-abundant regions. Recently, \citet{Narita_2024} also discusses the clumpy structures in the velocity components, which is consistent with our scenario. Secondly, it cannot be ruled out that this is due to the uncertainties in the reaction rate coefficients.

\subsection{Future Prospects}

In this work, we focus on carbon or oxygen-bearing molecules in the gas phase detected in diffuse or translucent clouds. We exclude nitrogen-bearing molecules from our discussion. The abundances of nitrogen-bearing molecules depend on the fractions of nitrogen atoms and N$_2$ since these two species are the major nitrogen reservoir in the gas phase. In order to estimate their fractions, we have to account for the self-shielding effect of N$_2$ \citep{Li_2013}, which depends on the spatial distribution of N$_2$ and thus requires implementation of a simplified nitrogen chemical network in the (magneto)hydrodynamics simulations. We will construct such a simple chemical network to be implemented in the hydrodynamics code in our future work.

The present work avoids discussion on the molecular ices other than H$_2$O ice. In the molecular cloud formation phase, the typical visual extinction is $A_V\sim 1\,\mathrm{mag}$. Then the dust temperature is higher than 10\,K \citep[e.g.][]{Hocuk_2017}, at which the thermal hopping of chemical species on the grain surface proceeds. Since the thermal hopping rate depends exponentially on the dust temperature, we need to treat the grain surface chemistry carefully, especially when we consider the time variation of physical conditions, including the dust temperature. Recently, \citet{Furuya_2024} developed a new astrochemical code including the binding energy distribution on the grain surface. One of the major effects of the binding energy distribution is to mitigate the dust temperature dependence of the thermal hopping rate. This may allow us to trace the formation history of molecular ices more stably.

\section{Summary}\label{sec:conclusion}

We investigated the molecular abundances of forming molecular clouds by an interstellar shock wave. We conducted 3D MHD simulations of the shock compression of atomic gas, including simple chemistry and tracer particles that move along the local velocity field. Then we performed detailed chemical network calculations along the trajectories of tracer particles as a post-process. We adopted three models to see the effects of physical history of fluid elements on the molecular abundances: the dynamical model, in which the time variation of physical parameters is considered, the static model, in which the physical parameters and the initial H$_2$ and H abundances are set to be the values of the final timestep of the dynamical model, and the H$_2$-fixed model, which is similar to the static model but the H$_2$ abundance is almost kept constant. Our findings are summarized as follows.

\begin{enumerate}
    \item H$_2$ exists both in diffuse and dense regions of the compression layer, whereas CO only exists in dense regions ($n_\mathrm{H}\gtrsim 10^3\,\mathrm{cm^{-3}}$) where the shielding of the external radiation field is sufficient. Thus, the diffuse region corresponds to "CO-poor" molecular clouds where the abundances of H$_2$ and CO are not linearly correlated.
    \item As the H$_2$ formation timescale is long and proportional to the inverse of gas density, the H$_2$ abundance depends on the density history of each fluid element, i.e., the duration of dense conditions. Therefore, fluid elements in diffuse regions can have abundant H$_2$ if they once stayed in high-density regions. 
    \item In diffuse conditions ($n_\mathrm{H}\lesssim 10^3\,\mathrm{cm^{-3}}$), the abundances of gas-phase molecules in the H$_2$-fixed model are consistent with those of the dynamical model, but those of the static model are not. This indicates that molecular abundances are mostly determined by steady-state chemistry in such diffuse regions once the H$_2$ abundance is given. 
    \item In dense conditions ($n_\mathrm{H}\gtrsim 10^3\,\mathrm{cm^{-3}}$), the molecular abundances of both the H$_2$-fixed and static models cannot reproduce those of the dynamical model, because the elemental C/O ratio in the gas phase sensitively reflects the ice formation along the flow. The boundary for the diffuse and dense conditions is determined by the threshold for H$_2$O ice formation on grain surfaces.
    \item The abundances of carbon-bearing molecules (CH and CCH) and oxygen-bearing molecules (OH and HCO$^+$) are determined by a quasi-steady-state chemistry. Our analytic solutions for the molecular abundances reproduce the results of the dynamical model unless the grain surface chemistry becomes important.
    \item We calculated the molecular column densities based on the spatial distribution of the tracer particles and their molecular abundances, and compared them with the diffuse molecular clouds. Although there are differences by factors of several to ten in the molecular abundances relative to H$_2$, our results can reproduce the linear correlations of column densities of CH, CCH, and OH with H$_2$ column density. Especially, CH and CCH show almost constant column density ratios relative to H$_2$, because the CH abundance saturates once H$_2$ formation is completed, and CCH is mainly formed from CH.
    \item Unlike the molecules above, HCO$^+$ shows non-linear dependence on H$_2$ column density, because HCO$^+$ abundance strongly depends on CO abundance. Our result at lower H$_2$ column density underestimates the typical abundance of HCO$^+$ derived from the observations of diffuse clouds. This may suggest that observed clouds are mixtures of CO-rich and CO-poor clouds.
\end{enumerate}

In the present study, we focused on carbon and oxygen-bearing molecules in the gas phase. We will extend the method adopted in this work and study the formation processes of nitrogen-bearing molecules and molecular ices other than H$_2$O in the cloud formation phase.

\section*{Acknowledgements}

We thank C.-C. Yang for providing the Lagrangian particle module of {\sc athena++}. We thank Shu-ichiro Inutsuka, Kengo Tomida, Nami Sakai, Joanne Dawson, Michael Busch, Daniel Rybarczyk, Masato Kobayashi, Shota Notsu, Yusuke Tsukamoto, Kanako Narita, Daniel Seifried, and Satoshi Ohashi for constructive discussions. The test calculations for the MHD simulations were conducted on Cray XC50 at the Center for Computational Astrophysics, National Astronomical Observatory in Japan. The numerical calculations for the MHD simulations were carried out on Yukawa-21 at YITP in Kyoto University. The post-process chemical network calculations were conducted on the general-purpose PC cluster at the Center for Computational Astrophysics, National Astronomical Observatory in Japan.
This work is supported by JSPS KAKENHI grant No. JP24KJ0901, JP21H04487, JP22KK0043, JP25K07364, JP20H05847, and JP24K00674. Y.K. acknowledges the support by International Graduate Program for Excellence in Earth-Space Science (IGPEES) of the University of Tokyo.

\section*{Data Availability}

The simulation data and the sample code for the analytic solution will be made available upon reasonable request.



\bibliographystyle{mnras}
\bibliography{citation} 




\appendix

\section{Reduced chemistry vs. detailed chemistry}\label{app:otf_ppc}

\begin{figure*}
	\includegraphics[width=2.0\columnwidth]{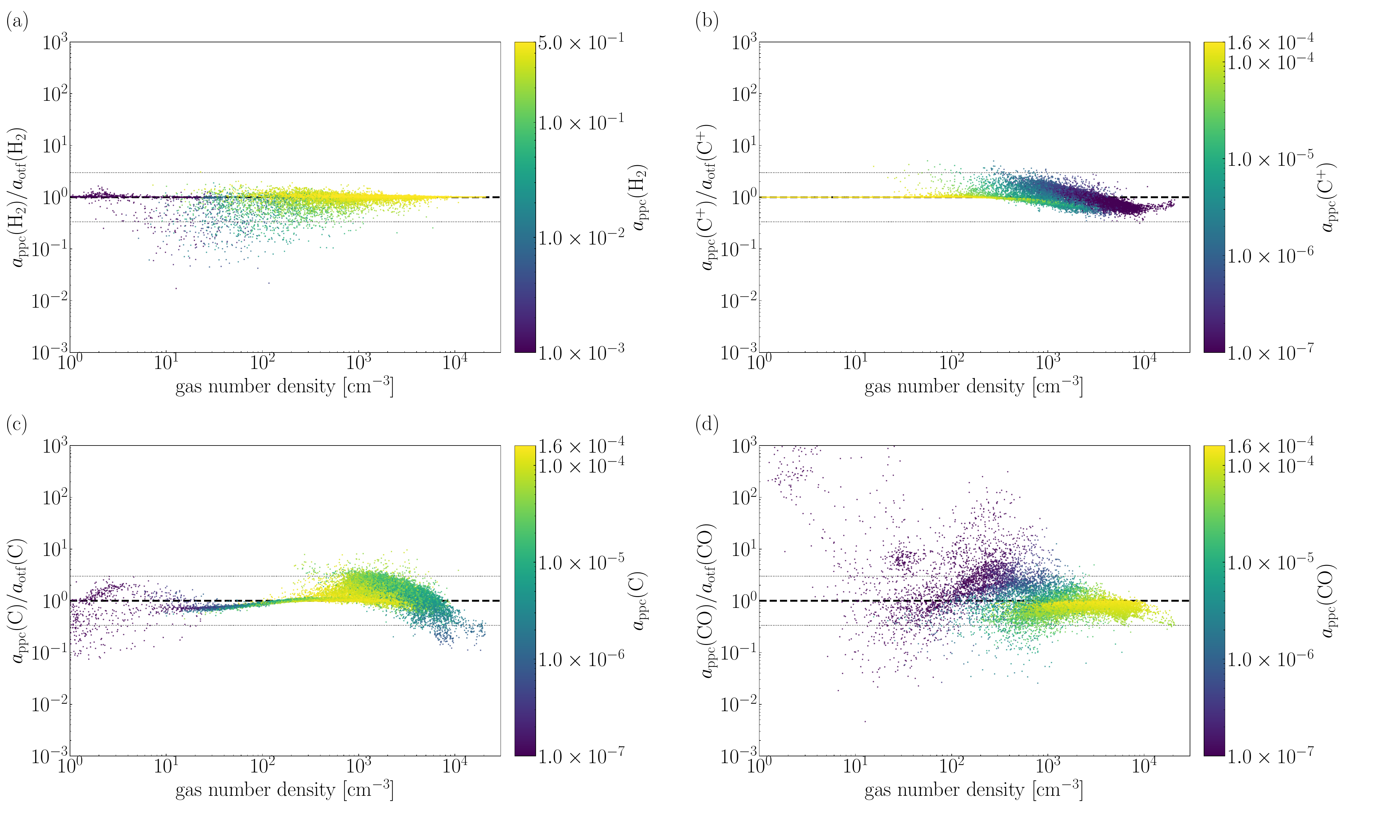}
    \caption{The gas number density and the ratio of abundances obtained from the post-process calculations ($a_\mathrm{ppc}$) to those from the small network calculations ($a_\mathrm{otf}$). The color scale shows the molecular abundances from the post-process calculations.}
    \label{fig:otf_vs_ppc}
\end{figure*}

In this work, we perform two chemical network calculations: small chemical network calculation implemented in MHD simulation and post-process calculation of large chemical network. Here we look into the consistency of the two calculations. The post-processing calculations were conducted with tracer particles, while the small network calculations were done in each cell. To compare the two results, we obtain the molecular abundances from the small network at the position of each tracer particle by interpolation (see Section \ref{sec:MHDsetup}).

Figure \ref{fig:otf_vs_ppc} shows the ratios of the abundances of major molecules obtained from the post-processing calculations ($a_\mathrm{ppc}$) against those from the small network calculations ($a_\mathrm{otf}$). The color scale shows $a_\mathrm{ppc}$. The thick dashed line shows $a_\mathrm{ppc}/a_\mathrm{otf}=1$, while the thin dotted lines show $a_\mathrm{ppc}/a_\mathrm{otf}=3\, \mathrm{and}\,1/3$. The ratio for H$_2$ is almost unity both in diffuse and dense regions (Panel (a)). Some particles show $a_\mathrm{ppc}/a_\mathrm{otf}<1/3$, which means that H$_2$ abundances are underestimated in the post-process calculations. The particles with low $a_\mathrm{ppc}(\mathrm{H_2}) (<10^{-1})$ tend to show this trend. This is mainly due to the different methods for the chemical network calculation in the MHD simulation and the post-process. 
The MHD simulation is performed by Euler method; the molecular abundances in each grid cell represents the averaged value of the (hypothetical) fluid elements that reached the cell via various trajectory. On the other hand, the post-process calculation, i.e., chemical network calculation along the trajectory of tracer particles, is a kind of Lagrange method. Consider as an example, a fluid percel that is injected during the calculation (Section \ref{sec:MHDsetup}). Since H$_2$ abundance sensitively depends on the density history (see Section \ref{sec:H2}), such "young" particles tend to have lower H$_2$ abundance compared with neighboring particles.
Then the H$_2$ abundance of "young" particles tends to be lower than that of grid cells where the abundance is averaged.
This kind of trend appears because the chemical timescale of H$_2$ is longer than the crossing timescale of fluid elements. The typical velocity dispersion in the compression layer is about $\sigma_v\sim 3\,\mathrm{km\,s^{-1}}$. The corresponding timescale for fluid elements to cross a clump with a size $l_\mathrm{clump}$ is about $0.3\,\mathrm{Myr}(l_\mathrm{clump}/1\,\mathrm{pc})(\sigma_v/3\,\mathrm{km\,s^{-1}})^{-1}$. The typical H$_2$ formation timescale is around $1-10\,\mathrm{Myr}$ at $n_\mathrm{H}\sim 10^2-10^3\,\mathrm{cm^{-3}}$, which is longer than the crossing timescale. 

On the other hand, $a_\mathrm{ppc}/a_\mathrm{otf}$ for C$^+$ and C at $n_\mathrm{H}<10^3\,\mathrm{cm^{-3}}$, where they are the major carbon carriers, has much smaller scatter than H$_2$. In this region, the photodissociation is effective. The chemical timescale for C$^+$ and C is determined by that of photodissociation, which is about $10^{-4}\,\mathrm{Myr}$ and much shorter than the crossing timescale. At $n_\mathrm{H}>10^3\,\mathrm{cm^{-3}}$, the abundances of C$^+$ and C are affected by CO abundance, which is the dominant carbon carrier there. But the scatter is about a factor of three, and the on-the-fly calculations and the post-process calculations show reasonable agreement.

Panel (d) shows the $a_\mathrm{ppc}/a_\mathrm{otf}$ for CO. The post-process calculations tend to show higher CO abundance than the on-the-fly calculations at $n_\mathrm{H}<10^3\,\mathrm{cm^{-3}}$. This is due to the difference in the chemical network. The large chemical network for the post process includes many species and reactions that are neglected in the small network. For instance, HOC$^+$ + e$^-$ $\rightarrow$ CO + H and CO$^+$ + H $\rightarrow$ H$^+$ + CO, which are not included in the small network, also contribute to CO formation in diffuse conditions. On the other hand, $a_\mathrm{ppc}/a_\mathrm{otf}$ is close to unity around $n_\mathrm{H}\gtrsim 10^3\,\mathrm{cm^{-3}}$. The results of the two calculations for CO reasonably agree when CO becomes abundant ($a_\mathrm{ppc}(\mathrm{CO})>1\times 10^{-6}$). $a_\mathrm{ppc}/a_\mathrm{otf}$ is slightly less than unity at $n_\mathrm{H}\sim 10^4\,\mathrm{cm^{-3}}$ because of the CO depletion, which is also not included in the reduced network.

\section{Effect of turbulence on H$_2$ formation}\label{app:turbulenceH2}

\begin{figure}
	\includegraphics[width=1.0\columnwidth]{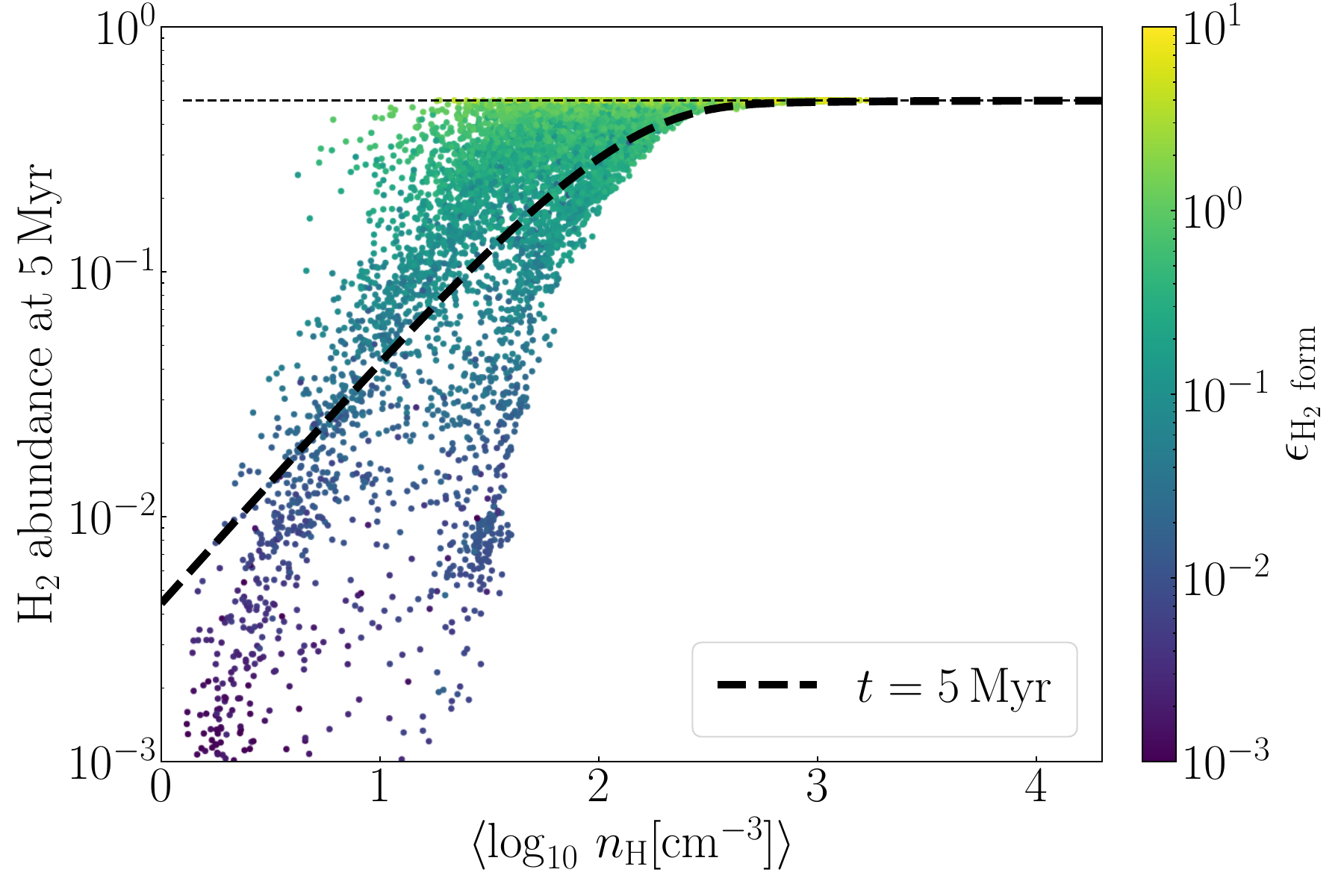}
    \caption{The average logarithmic gas number density and the H$_2$ abundance at 5 Myr of each tracer particles. The dashed line shows the analytic solution of H$_2$ abundance (equation (\ref{eq:H2analytic}))}, and the color scale shows the efficiency of H$_2$ formation, $\epsilon_\mathrm{H_2\ form}$ (equation (\ref{eq:H2formeff})).
    \label{fig:avenH_H2}
\end{figure}

We discuss the relationship between the density history of fluid elements and H$_2$ abundance in Section \ref{sec:H2}. Here we add notes about the effect of turbulence on H$_2$ formation. 
If the density enhancement due to turbulence plays an important role in H$_2$ formation, the H$_2$ abundances of fluid elements should higher than those predicted from equation (\ref{eq:H2analytic}) based on the average gas number densities experienced by the fluid elements. To verify this point, we calculate the average logarithmic gas number density for each tracer particle as
\begin{equation}
    \langle \log_{10}\,n_\mathrm{H}  \rangle=\frac{\sum_\mathrm{j}\log_{10}\,[n_\mathrm{H}(\mathrm{j})]\Delta t(\mathrm{j})}{\sum_\mathrm{j}\Delta t(\mathrm{j})}.
\end{equation}
Figure \ref{fig:avenH_H2} shows the averaged gas number density and the H$_2$ abundance at 5\,Myr of the tracer particles. The dashed line shows the analytic solution of H$_2$ abundance at 5\, Myr (equation \ref{eq:H2analytic}). Many tracer particles (about 87\% of tracer particles with $\langle \log_{10}\,n_\mathrm{H}  \rangle<3$) show higher H$_2$ abundances than the analytic solutions, which indicates that H$_2$ formation is accelerated in diffuse conditions due to a temporal increase in gas density by turbulence. It supports the conclusions of the previous studies: the rapid formation of H$_2$ through turbulent mixing \citep[e.g.][]{Glover_2007b}.

\section{Remapping from tracer particles}\label{app:remapping}

\begin{figure*}
	\includegraphics[width=1.8\columnwidth]{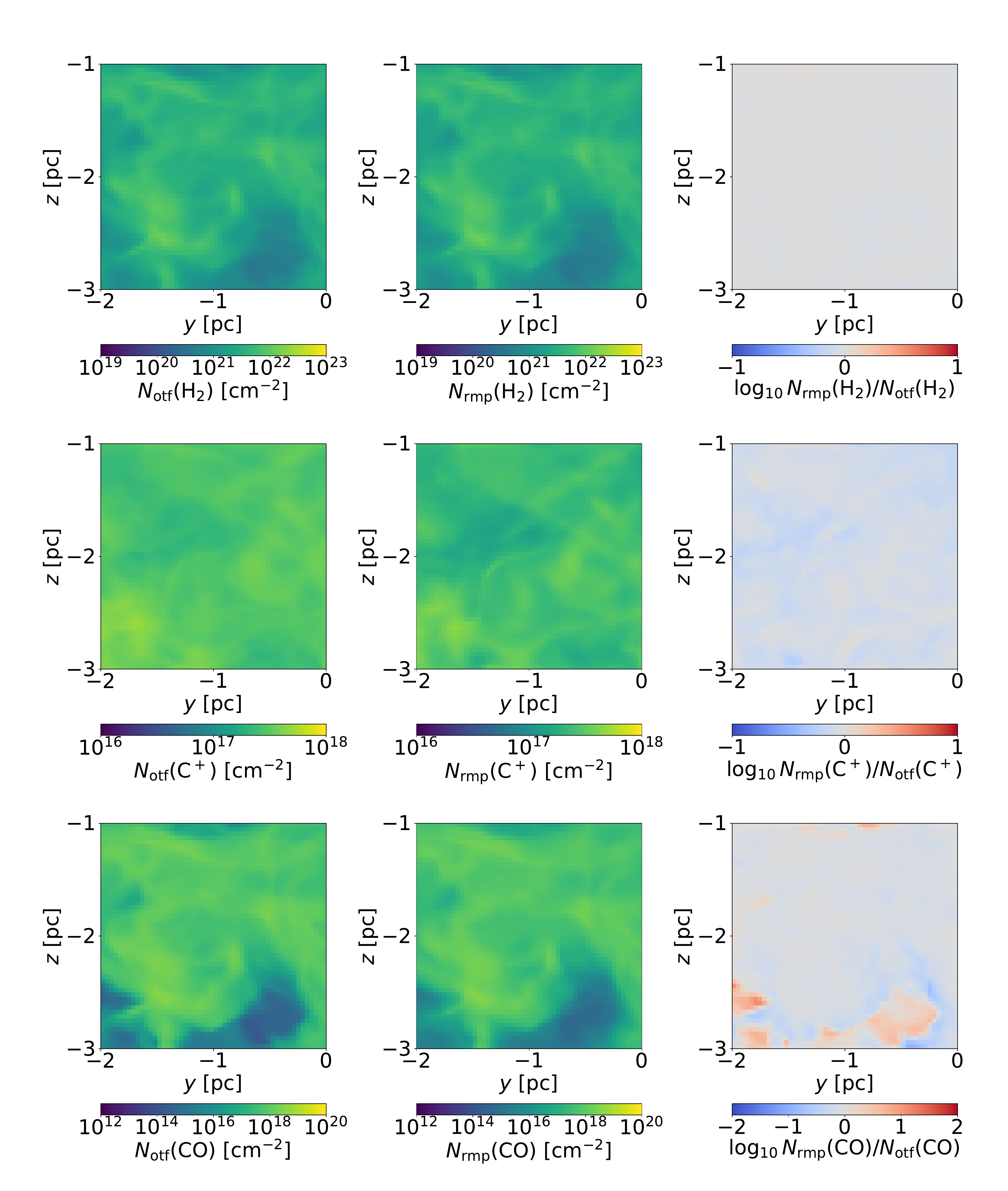}
    \caption{Left column: the H$_2$, C$^+$, and CO column densities from the on-the-fly calculations. Middle column: similar to the left column, but for the column densities from the remapping. Right column: the ratio of the column densities from the remapping against those from the on-the-fly calculations.}
    \label{fig:remap_compare}
\end{figure*}

\begin{figure}
	\includegraphics[width=1.0\columnwidth]{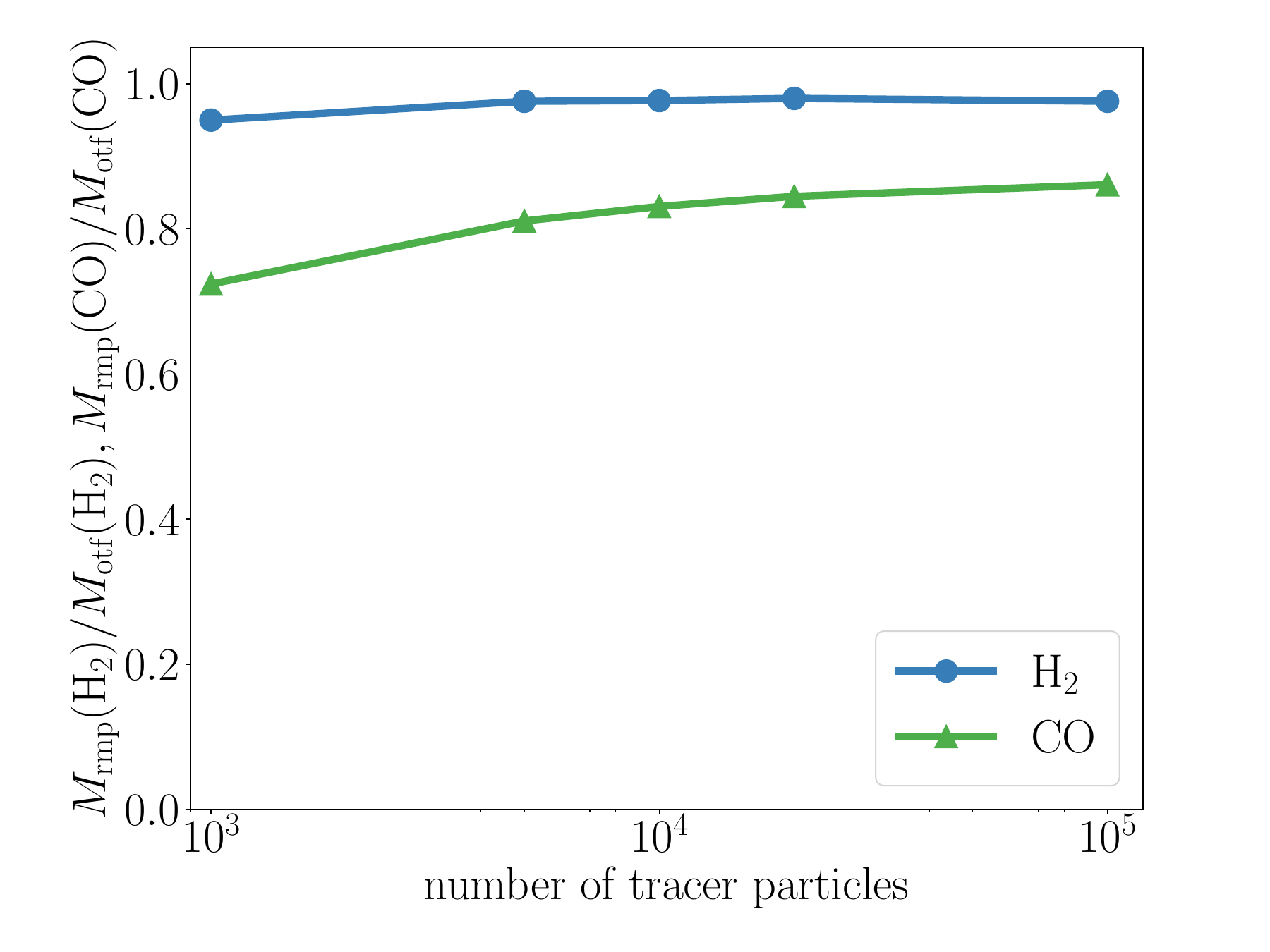}
    \caption{The ratio of H$_2$ (the blue circles) and CO (the green triangles) total mass from the remapping against those from the on-the-fly calculations.}
    \label{fig:remap_error}
\end{figure}

We derive molecular column densities from the spatial distribution of tracer particles and the results of the post-process chemical network calculations in Section \ref{sec:vs_obs}. Here we summarize the algorithm and the results of the test calculations to verify our method. The algorithm is similar to that often used for the visualization of the calculation results of the Smoothed Particle Hydrodynamics (SPH) simulations \citep[e.g.][]{Price_2007}. We assume the tracer particles to be smoothed particles, characterized by a volume defined with a kernel function and a smoothing length. Then we estimate molecular abundances at each grid cell by kernel interpolation.

The first step is to find neighbors for each tracer particle. We set the initial smoothing length to be 5\,pc so that the neighboring particles can be found in the pre-shock gas, where the number of tracer particles is small. Next, we obtain the smoothing length for each particle by iterating (a) and (b) described below, based on the variable smoothing length approach. 

(a) We calculate the "particle density" of each tracer particle $i$, $\rho_i^*$, as follows:
\begin{equation}
    \rho_i^*=\sum_j\,m_jW(|\vec{x_i}-\vec{x_j}|, C_\mathrm{smth}h_i),
\end{equation}
where $j$ represents the neighboring tracer particles for particle $i$, $m_j$ is a "mass" of each tracer particle, $W$ is a kernel function, $\vec{x}$ is the position of each tracer particle, $h_i$ is the smoothing length of particle $i$, and $C_\mathrm{smth}$ is a factor for smoothing \citep[see e.g.][]{Inutsuka_2002}. Here we define the neighboring tracer particles as those located within a radius of $3h_i$ from particle $i$. We adopt $W$ as 3D gaussian function, $W(r, h)=(\pi h^2)^{-3/2}\exp(-r^2/h^2)$, and $C_\mathrm{smth}=1.2$. Since the value of $m_j$ can be arbitrarily chosen for each particle, we set $m_j=1$ for simplicity. 

(b) We then obtain the new smoothing length as
\begin{equation}
    h_i=\eta \left(\frac{m_j}{\rho_i^*}\right)^{1/3},
\end{equation}
where $\eta$ is a constant. Here we set $\eta=1.2$. 

We continue this iteration until the smoothing length changes by less than 1\,\% of the previous one. Finally, we calculate the abundance of molecule A at grid cell k, $a_\mathrm{k}(\mathrm{A})$ as follows:
\begin{equation}
    \log(a_\mathrm{k}(\mathrm{A}))= \frac{\sum_\mathrm{i}\,(w_\mathrm{i}/\rho_\mathrm{i}^*)W(|\vec{x_\mathrm{i}}-\vec{x_\mathrm{k}}|, h_\mathrm{i})\log(a_\mathrm{i}(\mathrm{A}))}{\sum_\mathrm{i}\,(w_\mathrm{i}/\rho_\mathrm{i}^*)W(|\vec{x_\mathrm{i}}-\vec{x_\mathrm{k}}|, h_\mathrm{i})},
\end{equation}
where $w_i$ is a weight, $\vec{x_\mathrm{k}}$ is a position of grid cell k, and $a_i(\mathrm{A})$ is an abundance of A of tracer particle $i$. In this work, we set $w_\mathrm{i}=1$ for H$_2$ molecule, and $w_\mathrm{i}=a_\mathrm{i}(\mathrm{H_2})$ for other species. We confirmed that the remapping for CO abundance with $w_\mathrm{i}=a_\mathrm{i}(\mathrm{H_2})$ reproduces the original CO abundance distribution better than with $w_\mathrm{i}=1$.

We parallelize our code using the Framework for Developing Particle Simulator (FDPS, \citet{Iwasawa_2016}). To confirm that our algorithm described above reproduces the original distribution of molecular abundances correctly, we perform the remapping with $a_\mathrm{otf}$ and compare the results with MHD simulations. Figure \ref{fig:remap_compare} shows the column densities derived from the MHD simulations ($N_\mathrm{otf}$) for H$_2$, C$^+$, and CO (left column), the column densities from the remapping ($N_\mathrm{rmp}$, center column), and the ratio of the two results. H$_2$ column density distributions of the two results are very similar, and the ratio $N_\mathrm{rmp}/N_\mathrm{otf}$ is almost unity both in diffuse and dense regions. 

C$^+$ column densities also look similar, although the remapping results tend to underestimate the column densities. This may be due to the lack of tracer particles in diffuse regions. Since C$^+$ mainly exists at $n_\mathrm{H}\lesssim 10^2\,\mathrm{cm^{-3}}$ (see Figure \ref{fig:chemPDF}), the remapping results show larger errors compared with the case of H$_2$. But the difference is still within about a factor of two. 

Compared with H$_2$ and C$^+$, the remapping of CO abundances has larger errors, especially in the transition regions between diffuse and dense conditions. As described in Section \ref{sec:CO}, CO abundances sharply increase when the self and mutual shielding effects become effective. Therefore, there are some regions where the gradient of CO abundance is large. The mass fraction of such a region and the number of tracer particles in it are small, resulting in low spatial resolution. The remapped value is affected by CO abundance in denser regions. A similar situation is reported for remapping of vector quantities such as velocity \citep{Price_2007}. In other regions, the remapping results are reasonably consistent with the on-the-fly calculations.

Figure \ref{fig:remap_error} shows the ratio of the total mass of H$_2$ and CO from the remapping results ($M_\mathrm{rmp}(\mathrm{H_2}), M_\mathrm{rmp}(\mathrm{CO})$) to that of the on-the-fly calculations ($M_\mathrm{otf}(\mathrm{H_2}), M_\mathrm{otf}(\mathrm{CO})$) with different number of tracer particles. The total mass of H$_2$ already saturates and is almost the same as that of the on-the-fly calculations. The total mass of CO depends on the number of tracer particles. The ratio gradually increases as the number of tracer particles increases and is close to being saturated when the number is around $10^4$. The difference still exists, but it is only 15\,\%. Therefore, we conclude that our algorithm can reproduce the original molecular abundance distribution, and the error is small enough when we use 20,000 tracer particles for the remapping.

\section{Analytic solution of molecular abundances}\label{app:analyticsol}

\begin{figure*}
	\includegraphics[width=2.0\columnwidth]{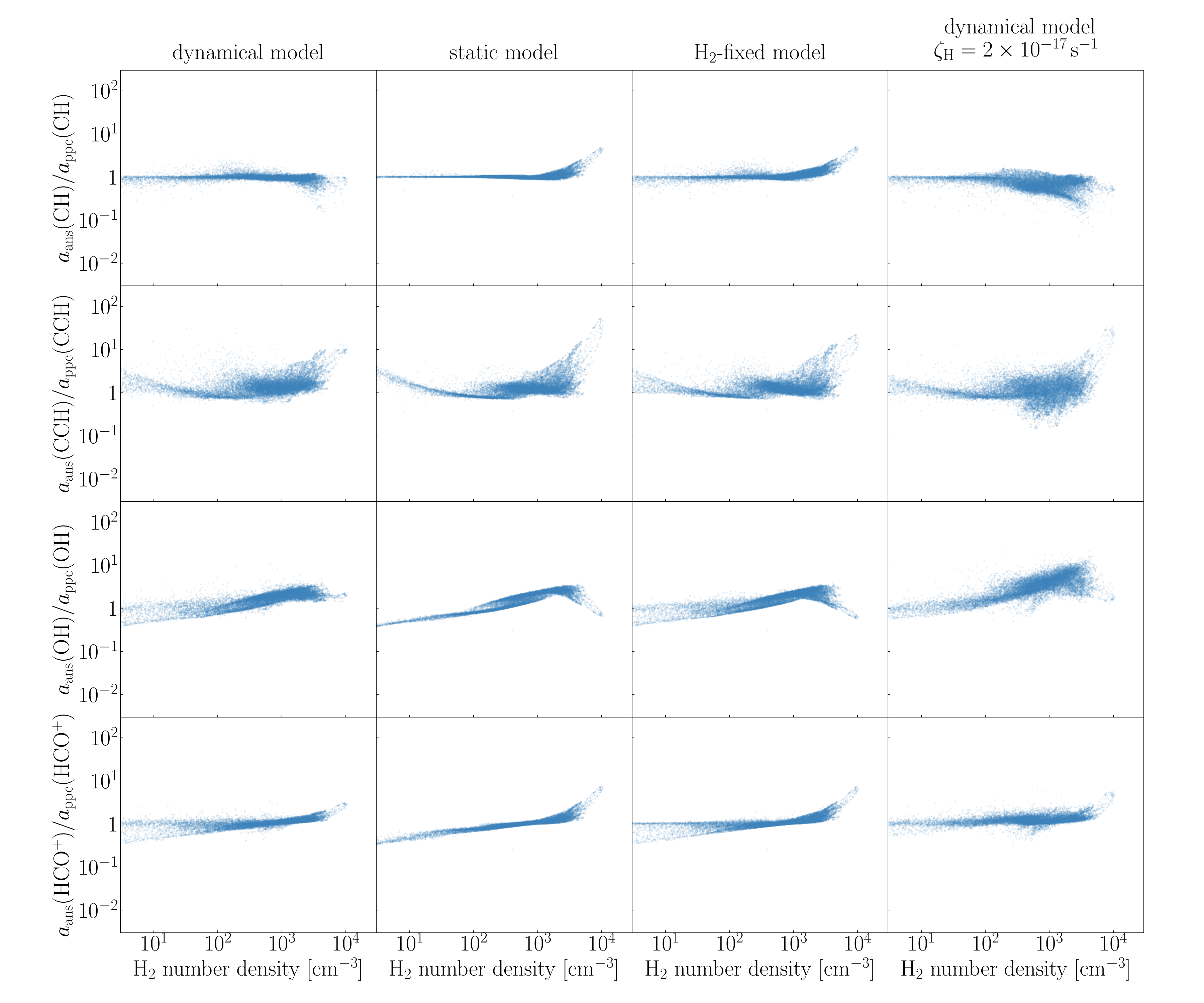}
    \caption{H$_2$ number density and the ratio of molecular abundances from the analytic solutions to those from the post-process calculations. Each column shows the results of the dynamical model, the static model, the H$_2$-fixed model, and the dynamical model with $\zeta_\mathrm{H}=2\times 10^{-17}\,\mathrm{s^{-1}}$ from left to right.}
    \label{fig:vs_analytic}
\end{figure*}

\begin{figure*}
	\includegraphics[width=1.8\columnwidth]{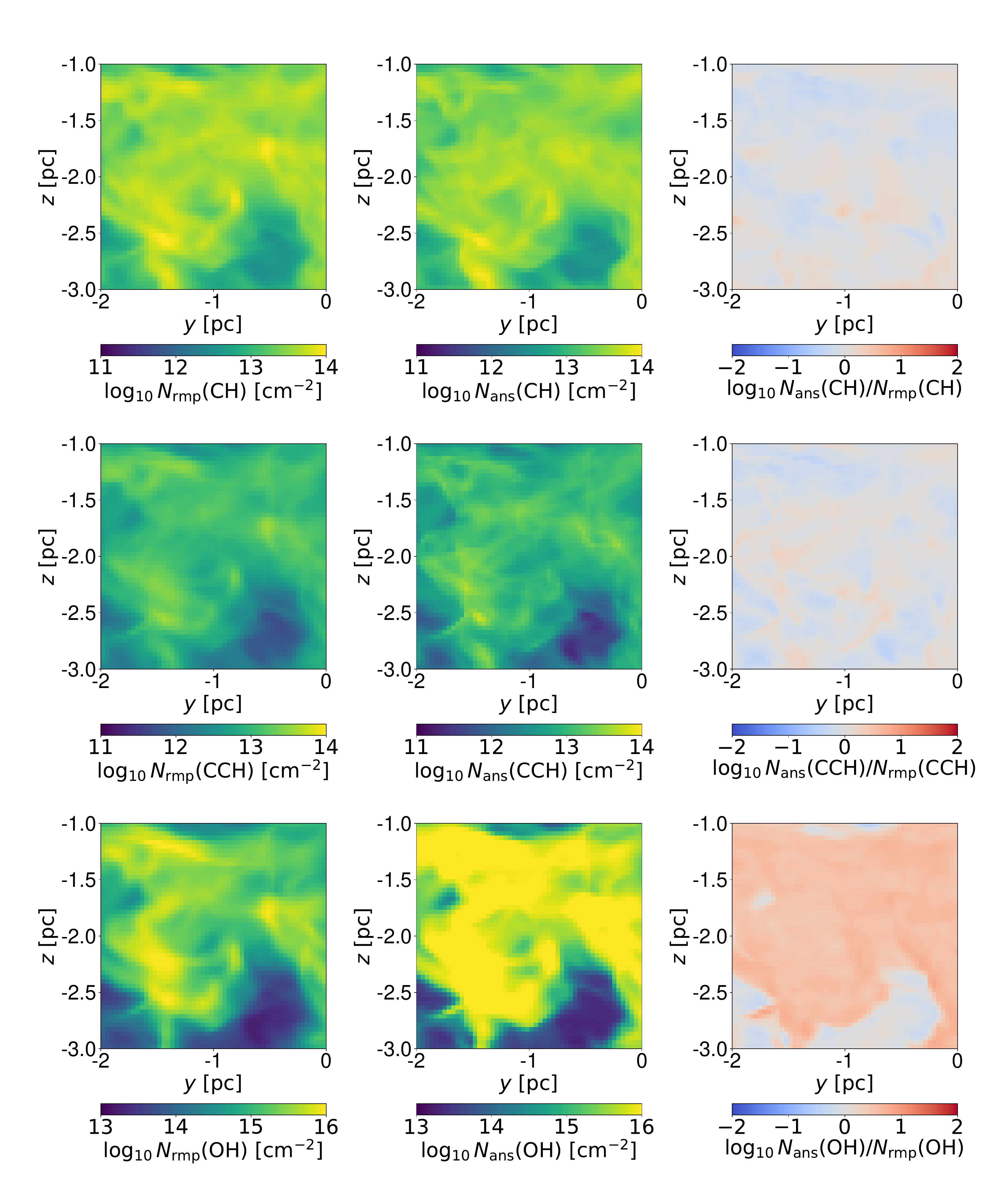}
    \caption{Left column: the CH, CCH, and OH column densities from the remapping. Middle column: similar to the left column, but for the column densities from the analytic solutions. Right column: the ratio of the column densities from the analytic solutions against those from the remapping.}
    \label{fig:remap_ans}
\end{figure*}

\begin{figure}
	\includegraphics[width=1.0\columnwidth]{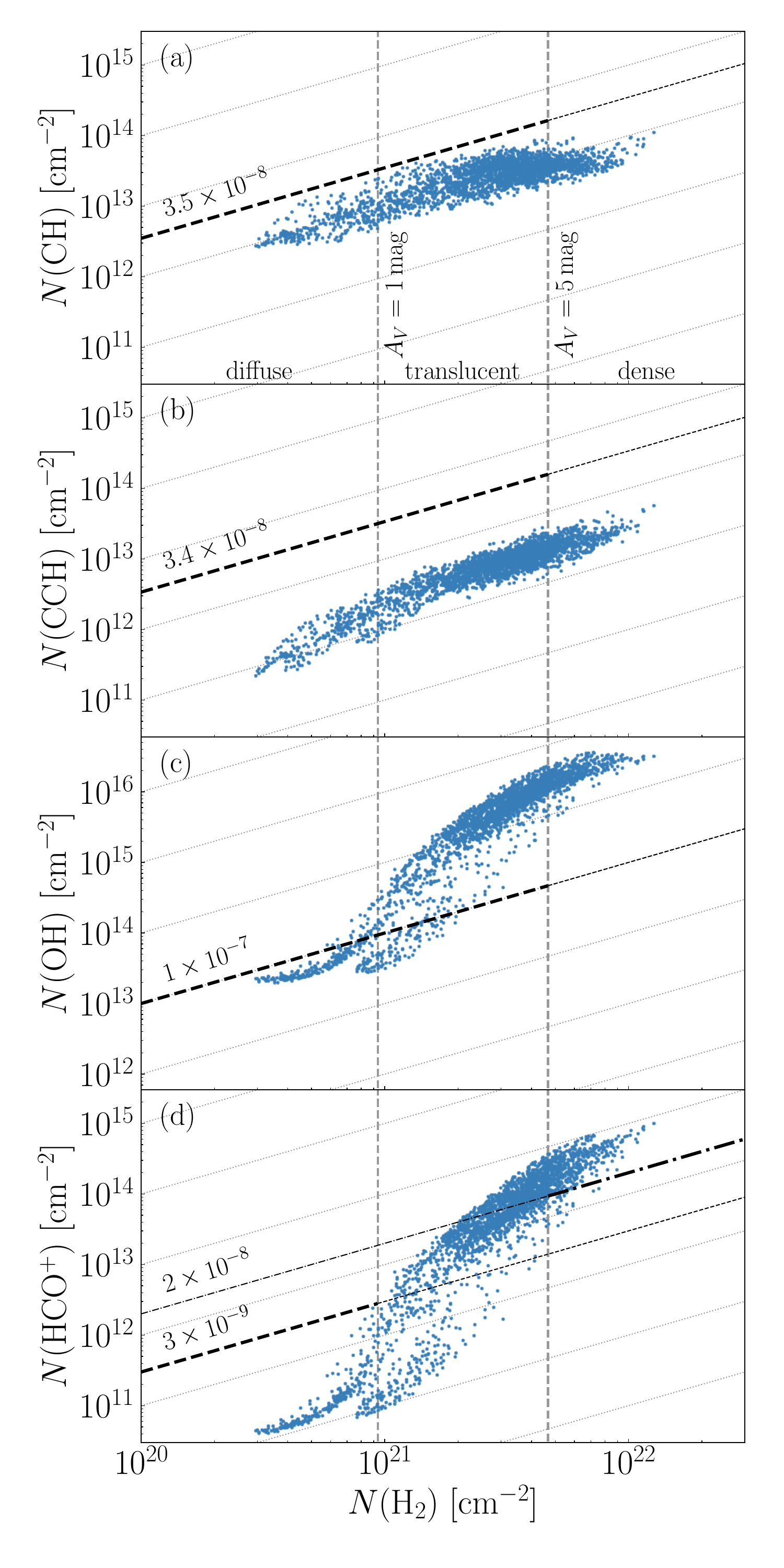}
    \caption{Similar to Figure \ref{fig:remap_mol}, but the blue dots show the calculated column densities from the analytic solutions.}
    \label{fig:remap_mol_ana}
\end{figure}

In this section, we describe the derivation of the analytic solution of the molecular abundances that we show in Section \ref{sec:each_molecule}. The formulation of the analytic solution is based on the steady-state assumption in the rate equations. Here, we define $k_\mathrm{A+B}$ as the rate coefficient for the reaction A + B, $f_\mathrm{A+B}^\mathrm{C}$ as the branching ratio of the reaction A + B $\rightarrow$ C + $\cdots$, $k_\mathrm{A,pd}$ as the photodissociation rate of A, and $k_\mathrm{A,dust}$ as the adsorption rate of A onto dust grains.

\subsection{CH and CCH}

In steady-state, the balance between major formation and destruction reactions of CH, CH$_2$, CH$^+$, CH$_2^+$, and CH$_3^+$ is given as
\begin{equation}\label{eq:CH}
    \begin{split}
         &k_\mathrm{CH_2+H}n(\mathrm{CH_2})n(\mathrm{H})+f_\mathrm{CH_3^++e^-}^\mathrm{CH}k_\mathrm{CH_3^++e^-}n(\mathrm{CH_3^+})n(\mathrm{e^-})\\&+f_\mathrm{CH_2^++e^-}^\mathrm{CH}k_\mathrm{CH_2^++e^-}n(\mathrm{CH_2^+})n(\mathrm{e^-})\\&-k_\mathrm{CH+H}n(\mathrm{CH})n(\mathrm{H})-k_\mathrm{CH,pd}n(\mathrm{CH})=0,
    \end{split}
\end{equation}
\begin{equation}\label{eq:CH2}
    f_\mathrm{CH_3^++e^-}^\mathrm{CH_2}k_\mathrm{CH_3^++e^-}n(\mathrm{CH_3^+})n(\mathrm{e^-})-k_\mathrm{CH_2+H}n(\mathrm{CH_2})n(\mathrm{H})=0,
\end{equation}
\begin{equation}\label{eq:CHp}
    k_\mathrm{C+H_3^+}n(\mathrm{C})n(\mathrm{H_3^+})-k_\mathrm{CH^++H_2}n(\mathrm{CH^+})n(\mathrm{H_2})=0,
\end{equation}
\begin{equation}\label{eq:CH2p}
    \begin{split}
        &k_\mathrm{C^++H_2}n(\mathrm{C^+})n(\mathrm{H_2})+k_\mathrm{CH^++H_2}n(\mathrm{CH^+})n(\mathrm{H_2})\\&-k_\mathrm{CH_2^++H_2}n(\mathrm{CH_2^+})n(\mathrm{H_2})-k_\mathrm{CH_2^++e^-}n(\mathrm{CH_2^+})n(\mathrm{e^-})=0,
    \end{split}
\end{equation}
and,
\begin{equation}\label{eq:CH3p}
    k_\mathrm{CH_2^++H_2}n(\mathrm{CH_2^+})n(\mathrm{H_2})-k_\mathrm{CH_3^++e^-}n(\mathrm{CH_3^+})n(\mathrm{e^-})=0.
\end{equation}
Combining the equations (\ref{eq:CH})-(\ref{eq:CH3p}), we can derive the number density of CH with equation (\ref{eq:nCH}) in Section \ref{sec:ans_ab}.

To estimate the molecular abundance of CCH, first we derive the steady-state solution of C$_2$H$_2^+$ abundance from
\begin{equation}\label{eq:C2H2p}
    \begin{split}
         &k_\mathrm{C_2H^++H_2}n(\mathrm{C_2H^+})n(\mathrm{H_2})+k_\mathrm{CH_3^++C}n(\mathrm{CH_3^+})n(\mathrm{C})\\&+f_\mathrm{CH_4+C^+}^\mathrm{C_2H_2^+}k_\mathrm{CH_4+C^+}n(\mathrm{CH_4})n(\mathrm{C^+})\\&-k_\mathrm{C_2H_2^++e^-}n(\mathrm{C_2H_2^+})n(\mathrm{e^-})-k_\mathrm{C_2H_2^++H_2}n(\mathrm{C_2H_2^+})n(\mathrm{H_2})=0.
    \end{split}
\end{equation}
The number density of CH$_3^+$ is derived from equations (\ref{eq:CHp}) - (\ref{eq:CH3p}). The C$_2$H$^+$ number density is given by the steady-state equation of C$_2^+$ and C$_2$H$^+$:
\begin{equation}\label{eq:C2p}
    k_\mathrm{C^++CH}n(\mathrm{C^+})n(\mathrm{CH})-k_\mathrm{C_2^++H_2}n(\mathrm{C_2^+})n(\mathrm{H_2})=0,
\end{equation}
and
\begin{equation}\label{eq:C2Hp}
    k_\mathrm{C_2^++H_2}n(\mathrm{C_2^+})n(\mathrm{H_2})-k_\mathrm{C_2H^++H_2}n(\mathrm{C_2H^+})n(\mathrm{H_2})=0.
\end{equation}
The number density of CH$_4$ is related to grain surface chemistry; The hydrogenation of atomic carbon and the desorption of CH$_4$ ice contributes the formation of gas-phase CH$_4$. Here we simply assume that a fraction $p_\mathrm{C}$ of the adsorbed atomic carbon is converted to gas-phase CH$_4$. Then the steady state is given as
\begin{equation}\label{eq:CH4}
    p_\mathrm{C}k_\mathrm{C, dust}n(\mathrm{C})-k_\mathrm{CH_4,pd}n(\mathrm{CH_4})-k_\mathrm{CH_4+C^+}n(\mathrm{CH_4})n(\mathrm{C^+})=0.
\end{equation}
Beside C$_2$H$_2^+$, C$_2$H$_4^+$, C$_2$H$_3^+$, and C$_2$H$_2$ also help CCH formation. Their steady state equations are as follows:
\begin{equation}\label{eq:C2H4p}
    \begin{split}
        &k_\mathrm{C_2H_2^++H_2}n(\mathrm{C_2H_2^+})n(\mathrm{H_2})-k_\mathrm{C_2H_4^++H}n(\mathrm{C_2H_4^+})n(\mathrm{H})\\&-k_\mathrm{C_2H_4^++e^-}n(\mathrm{C_2H_4^+})n(\mathrm{e^-})=0,
    \end{split}
\end{equation}
\begin{equation}\label{eq:C2H3p}
    \begin{split}
        &f_\mathrm{CH_4+C^+}^\mathrm{C_2H_3^+}k_\mathrm{CH_4+C^+}n(\mathrm{CH_4})n(\mathrm{C^+})\\&+k_\mathrm{C_2H_4^++H}n(\mathrm{C_2H_4^+})n(\mathrm{H})-k_\mathrm{C_2H_3^++e^-}n(\mathrm{C_2H_3^+})n(\mathrm{e^-})=0,
    \end{split}
\end{equation}
and,
\begin{equation}\label{eq:C2H2}
    \begin{split}
        &f_\mathrm{C_2H_4^++e^-}^\mathrm{C_2H_2}k_\mathrm{C_2H_4^++e^-}n(\mathrm{C_2H_4^+})n(\mathrm{e^-})\\&+f_\mathrm{C_2H_3^++e^-}^\mathrm{C_2H_2}k_\mathrm{C_2H_3^++e^-}n(\mathrm{C_2H_3^+})n(\mathrm{e^-})\\&-k_\mathrm{C_2H_2+C}n(\mathrm{C_2H_2})n(\mathrm{C})-k_\mathrm{C_2H_2,pd}n(\mathrm{C_2H_2})=0.
    \end{split}
\end{equation}
Finally, the steady-state abundance of CCH is given as
\begin{equation}\label{eq:CCH}
    \begin{split}
        &f_\mathrm{C_2H_2^++e^-}^\mathrm{CCH}k_\mathrm{C_2H_2^++e^-}n(\mathrm{C_2H_2^+})n(\mathrm{e^-})\\&+f_\mathrm{C_2H_3^++e^-}^\mathrm{CCH}k_\mathrm{C_2H_3^++e^-}n(\mathrm{C_2H_3^+})n(\mathrm{e^-})\\&+k_\mathrm{C_2H_2,pd}n(\mathrm{C_2H_2})-k_\mathrm{CCH+C^+}n(\mathrm{CCH})n(\mathrm{C^+})\\&-k_\mathrm{CCH+O}n(\mathrm{CCH})n(\mathrm{O})-k_\mathrm{CCH,pd}n(\mathrm{CCH})=0.
    \end{split}
\end{equation}
Equation (\ref{eq:CCH}) can be solved for the number density of CCH as equation (\ref{eq:nCCH}).

Equation \ref{eq:nCH} includes H$_3^+$ number density, which can be estimated from the steady-state balance for H$_2^+$ and H$_3^+$:
\begin{equation}\label{eq:H2p}
    \zeta_\mathrm{cr, H_2}n(\mathrm{H_2})-k_\mathrm{H_2^++H_2}n(\mathrm{H_2^+})n(\mathrm{H_2})-k_\mathrm{H_2^++e^-}n(\mathrm{H_2^+})n(\mathrm{e^-})=0,
\end{equation}
and,
\begin{equation}\label{eq:H3p}
    \begin{split}
        &k_\mathrm{H_2^++H_2}n(\mathrm{H_2^+})n(\mathrm{H_2})-k_\mathrm{H_3^++e^-}n(\mathrm{H_3^+})n(\mathrm{e^-})-k_\mathrm{H_3^++C}n(\mathrm{H_3^+})n(\mathrm{C})\\&-k_\mathrm{H_3^++O}n(\mathrm{H_3^+})n(\mathrm{O})-k_\mathrm{H_3^++CO}n(\mathrm{H_3^+})n(\mathrm{CO})=0.
    \end{split}
\end{equation}
Combining the equations (\ref{eq:CH}) - (\ref{eq:H3p}), the number densities of CH and CCH can be expressed as a function of the number density of major species (H, H$_2$, C$^+$, C, O, CO, and e$^-$) and the physical quantities (gas number density, gas temperature, and visual extinction).

\subsection{OH and HCO$^+$}

OH formation starts from OH$^+$. The steady-state equation is
\begin{equation}\label{eq:OHp}
    \begin{split}
        &k_\mathrm{O^++H_2}n(\mathrm{O^+})n(\mathrm{H_2})+f_\mathrm{H_3^++O}^\mathrm{OH^+}k_\mathrm{H_3^++O}n(\mathrm{H_3^+})n(\mathrm{O})\\&-k_\mathrm{OH^++H_2}n(\mathrm{OH^+})n(\mathrm{H_2})=0.
    \end{split}
\end{equation}
The number density of O$^+$ can be estimated by the following steady-state equation using the number densities of H, H$_2$, H$^+$, and O:
\begin{equation}\label{eq:Op}
    \begin{split}
        &\zeta_\mathrm{cr,O}n(\mathrm{O})+k_\mathrm{H^++O}n(\mathrm{H^+})n(\mathrm{O})\\&-k_\mathrm{O^++H_2}n(\mathrm{O^+})n(\mathrm{H_2})-k_\mathrm{O^++H}n(\mathrm{O^+})n(\mathrm{H})=0,
    \end{split}
\end{equation}
where $\zeta_\mathrm{cr,O}$ is the cosmic ray ionization rate of an oxygen atom. OH is formed by the series of reactions with H$_2$ and the electron recombination reactions. The steady-state equations of H$_2$O$^+$, H$_3$O$^+$, and OH are given as follows:
\begin{equation}\label{eq:H2Op}
    \begin{split}
        &k_\mathrm{OH^++H_2}n(\mathrm{OH^+})n(\mathrm{H_2})+f_\mathrm{H_3^++O}^\mathrm{H_2O^+}k_\mathrm{H_3^++O}n(\mathrm{H_3^+})n(\mathrm{O})\\&+k_\mathrm{H_3^++OH}n(\mathrm{H_3^+})n(\mathrm{OH})\\&-k_\mathrm{H_2O^++H_2}n(\mathrm{H_2O^+})n(\mathrm{H_2})-k_\mathrm{H_2O^++e^-}n(\mathrm{H_2O^+})n(\mathrm{e^-})=0,
    \end{split}
\end{equation}
\begin{equation}\label{eq:H3Op}
    k_\mathrm{H_2O^++H_2}n(\mathrm{H_2O^+})n(\mathrm{H_2})-k_\mathrm{H_3O^++e^-}n(\mathrm{H_3O^+})n(\mathrm{e^-})=0,
\end{equation}
and,
\begin{equation}\label{eq:OH}
    \begin{split}
        &f_\mathrm{H_3O^++e^-}^\mathrm{OH}k_\mathrm{H_3O^++e^-}n(\mathrm{H_3O^+})n(\mathrm{e^-})+p_\mathrm{O}k_\mathrm{O, dust}n(\mathrm{O})\\&-k_\mathrm{C^++OH}n(\mathrm{C^+})n(\mathrm{OH})-k_\mathrm{H^++OH}n(\mathrm{H^+})n(\mathrm{OH})\\&-k_\mathrm{O+OH}n(\mathrm{O})n(\mathrm{OH})-k_\mathrm{S+OH}n(\mathrm{S})n(\mathrm{OH})\\&-k_\mathrm{H_3^++OH}n(\mathrm{H_3^+})n(\mathrm{OH})-k_\mathrm{OH,pd}n(\mathrm{OH})=0.
    \end{split}
\end{equation}
The second term of equation (\ref{eq:OH}) represents the contribution from the grain surface chemistry. Similar to the case of CH$_4$, we simply assume that a fraction $p_\mathrm{O}$ of the adsorbed oxygen atoms are converted to gas-phase OH. Using equations (\ref{eq:OHp}) - (\ref{eq:H3Op}) and the number density of H$_3^+$, the OH number density is given as equation (\ref{eq:nOH}).
Here we use $f_\mathrm{H_3^++O}^\mathrm{OH^+}+f_\mathrm{H_3^++O}^\mathrm{H_2O^+}=1$.

There are three pathways for HCO$^+$ formation. One is from OH and CO$^+$. The steady-state equation of CO$^+$ is
\begin{equation}\label{eq:COp}
    \begin{split}
        &k_\mathrm{C^++OH}n(\mathrm{C^+})n(\mathrm{OH})-k_\mathrm{CO^++H_2}n(\mathrm{CO^+})n(\mathrm{H_2})\\&-k_\mathrm{CO^++H}n(\mathrm{CO^+})n(\mathrm{H})-k_\mathrm{CO^++e^-}n(\mathrm{CO^+})n(\mathrm{e^-})=0.
    \end{split}
\end{equation}
Therefore, CO$^+$ number density is directly derived from OH number density. The second path is the two-body reaction of H$_3^+$ + CO $\rightarrow$ HCO$^+$ + H$_2$. The third one is from H$_2$O in the gas phase. The steady-state equation of H$_2$O is given as
\begin{equation}\label{eq:H2O}
    \begin{split}
        &p_\mathrm{O}k_\mathrm{O, dust}n(\mathrm{O})-k_\mathrm{C^++H_2O}n(\mathrm{C^+})n(\mathrm{H_2O})\\&-k_\mathrm{H^++H_2O}n(\mathrm{H^+})n(\mathrm{H_2O})-k_\mathrm{H_2O,pd}n(\mathrm{H_2O})=0.
    \end{split}
\end{equation}
The first term represents the desorption of H$_2$O from grain surfaces. Then the steady-state equation of HCO$^+$ is given as
\begin{equation}\label{eq:HCOp}
    \begin{split}
        &f_\mathrm{CO^++H_2}^\mathrm{HCO^+}k_\mathrm{CO^++H_2}n(\mathrm{CO^+})n(\mathrm{H_2})+k_\mathrm{H_3^++CO}n(\mathrm{H_3^+})n(\mathrm{CO})\\&+f_\mathrm{H_3^++CO}^\mathrm{HCO^+}k_\mathrm{C^++H_2O}n(\mathrm{C^+})n(\mathrm{H_2O})-k_\mathrm{HCO^++e^-}n(\mathrm{HCO^+})n(\mathrm{e^-})=0.
    \end{split}
\end{equation}
Combining equations (\ref{eq:H2p}) - (\ref{eq:HCOp}), the number density of HCO$^+$ can be expressed as a function of the number densities of H, H$_2$, H$^+$, C$^+$, C, O, CO, S, and e$^-$. The reaction S + OH $\rightarrow$ SO + H becomes important only if gas density is large enough and sulfur atoms exist as S. Therefore, we set the number density of S by the elemental abundance (Table \ref{tab:InitialAbundance}).

\subsection{Application of the analytic solutions}

We calculate the molecular abundances using the analytic solutions ($a_\mathrm{ans}$) described above. We set $p_\mathrm{C}=1$ and $p_\mathrm{O}=1$ for simplicity. Figure \ref{fig:vs_analytic} shows the ratio of $a_\mathrm{ans}$ against $a_\mathrm{ppc}$ for the dynamical model (the column on the far left), the static model (second from the left), and the H$_2$-fixed model (third from the left). The ratio $a_\mathrm{ans}/a_\mathrm{ppc}$ is almost unity with a scatter of a factor of a few at $n_\mathrm{H}<10^3\,\mathrm{cm^{-3}}$ for all models. On the other hand, the ratio deviates from unity at $n_\mathrm{H}\sim 10^4\,\mathrm{cm^{-3}}$ in the dynamical, static, and H$_2$-fixed models. The reason for this trend is that the major formation paths change from those we use in the analytic solution due to the grain surface chemistry. In the case of CCH, the ratio $a_\mathrm{ans}/a_\mathrm{ppc}$ exceed unity at $n_\mathrm{H}>10^3\,\mathrm{cm^{-3}}$ because of the high value of $p_\mathrm{C}$. In such a dense condition, the conversion of adsorbed carbon atoms to CH$_4$ is not so efficient due to the long timescale of photodesorption compared to the adsorption timescale (see also Section \ref{sec:dynamics_static}).

The threshold for the applicability of our analytic solution is roughly determined by the competition between ion-neutral reactions and grain surface reactions. The typical timescale of the ion-neutral reaction $\tau_\mathrm{in}$ is given by the Langevin rate coefficient $k_\mathrm{L}$ and the number density of H$_3^+$. In dense condition, H$_3^+$ is mainly destroyed by CO. From equations (\ref{eq:H2p}) and (\ref{eq:H3p}), H$_3^+$ number density is roughly given by 
\begin{equation}\label{eq:nH3p}
    n(\mathrm{H_3^+})\simeq \frac{\zeta_\mathrm{H_2}n(\mathrm{H_2})}{k_\mathrm{L}n(\mathrm{CO})}.
\end{equation}
Then $\tau_\mathrm{in}$ is expressed as
\begin{equation}
    \begin{split}
        \tau_\mathrm{in}&=\frac{1}{k_\mathrm{L}n(\mathrm{H_3^+})}\simeq \frac{n(\mathrm{CO})}{\zeta_\mathrm{H_2}n(\mathrm{H_2})}\\&\sim 10^{12}\,\mathrm{s}\left(\frac{\zeta_\mathrm{H_2}}{10^{-16}\,\mathrm{s^{-1}}}\right)^{-1}.
    \end{split}
\end{equation}
The timescale of grain surface chemistry is represented by the timescale of the adsorption of atoms to dust grains. For example, the timescale for an oxygen atom, $\tau_\mathrm{ad}(\mathrm{O})$, is described as follows:
\begin{equation}
    \tau_\mathrm{ad}(\mathrm{O})=\frac{1}{\sigma_\mathrm{gr}v_\mathrm{th}(\mathrm{O})n_\mathrm{gr}}\sim 10^{12}\,\mathrm{s} \left(\frac{T}{10\,\mathrm{K}}\right)^{-0.5}\left(\frac{n_\mathrm{H}}{10^5\,\mathrm{cm^{-3}}}\right)^{-1},
\end{equation}
where $v_\mathrm{th}(\mathrm{O})$ represents the thermal velocity of an oxygen atom. Then the ratio of the two timescales is expressed as
\begin{equation}\label{eq:ad_vs_in}
    \frac{\tau_\mathrm{ad}(\mathrm{O})}{\tau_\mathrm{in}}\sim1\left(\frac{\zeta_\mathrm{H_2}}{10^{-16}\,\mathrm{s^{-1}}}\right)\left(\frac{T}{10\,\mathrm{K}}\right)^{-0.5}\left(\frac{n_\mathrm{H}}{10^5\,\mathrm{cm^{-3}}}\right)^{-1}.
\end{equation}
If gas density is much smaller than about $10^{5}\,\mathrm{cm^{-3}}$, the adsorption timescale is shorter than the timescale of the ion-neutral reaction, and our analytic solution can be applicable. Note that H$_3^+$ number density given by (\ref{eq:nH3p}) is a kind of upper limit value, and the threshold gas density can be smaller than $10^5\,\mathrm{cm^{-3}}$ by a factor of three. 

Equation (\ref{eq:ad_vs_in}) indicates that the threshold gas density decreases when the cosmic-ray ionization rate is lower. To verify this point, we conduct an additional post-process calculation. The setup is almost identical to that of the dynamical model, but the cosmic-ray ionization rate is set to $\zeta_\mathrm{H}=2\times 10^{-17}\,\mathrm{s^{-1}}$, which is ten times smaller than the fiducial value. The result is shown in the column on the far right in Figure \ref{fig:vs_analytic}. As expected from equation (\ref{eq:ad_vs_in}), the ratio $a_\mathrm{ans}/a_\mathrm{ppc}$ tends to deviate from unity at a lower gas density ($n_\mathrm{H}\sim 10^3\,\mathrm{cm^{-3}}$) than in the dynamical model with the fiducal ionization rate ($n_\mathrm{H}\sim 10^4\,\mathrm{cm^{-3}}$). 

Since our analytic solutions only use the abundances of major chemical species that are included in the simple chemical network, we can use them to calculate molecular abundances in all grid cells of our MHD calculations. For example, Figure \ref{fig:remap_ans} shows the column densitiy maps of CH, CCH, and OH calculated by the remapping (left column) and by the analytic solutions (middle column), while the right column shows their ratio. Our analytic solutions reasonably reproduce the column density maps derived from the remapping. Figure \ref{fig:remap_mol_ana} shows the column densities of H$_2$ and the molecules discussed in Section \ref{sec:vs_obs}. The analytic solutions also reproduce the relationship between H$_2$ and the molecular column densities from the remapping (Figure \ref{fig:remap_mol}). In the case of OH, the column densities from the analytic solutions are about three times higher than those from the remapping in higher H$_2$ column density regions. This discrepancy arises from the absence of H$_2$O ice formation in the small network implemented in our MHD simulation.


\bsp	
\label{lastpage}
\end{document}